\documentclass[%
 reprint,
superscriptaddress,
 amsmath,amssymb,
 aps,pre,hyphens,floatfix
]{revtex4-2}

\usepackage[]{amsmath}
\usepackage{amssymb}
\usepackage{enumerate}
\usepackage[usenames,dvipsnames,svgnames,table]{xcolor}
\usepackage{IEEEtrantools}
\usepackage{graphicx}% Include figure files
\usepackage{array}
\usepackage{multirow}
\usepackage{wrapfig}
\usepackage{fullpage}
\usepackage{overpic}
\usepackage{verbatim}
\usepackage{float}
\usepackage{neuralnetwork}
\usepackage[caption=false]{subfig}
\usepackage{multirow}
\usepackage{diagbox}

\usepackage[normalem]{ulem}  % without normalem the long lines in the bibliography do not break up!

\begin{document}

\preprint{APS/123-QED}

\title{Supervised and unsupervised learning of directed percolation}
% Force line breaks with \\
%\author{Complexity Science}
%%\email[]{sjm@mails.ccnu.edu.com}
%\affiliation{Key Laboratory of Quark and Lepton Physics (MOE) and Institute of Particle Physics, Central China Normal University, Wuhan 430079, China}

\author{Jianmin Shen}
%\email[]{sjm@mails.ccnu.edu.com}
\affiliation{Key Laboratory of Quark and Lepton Physics (MOE) and Institute of Particle Physics, Central China Normal University, Wuhan 430079, China}
%\affiliation{Department of Physics and Center for Soft Matter and Biological Physics, \\
%	Virginia Tech, Blacksburg, VA 24061, USA}
\author{Wei Li}
\email[]{liw@mail.ccnu.edu.cn}
\affiliation{Key Laboratory of Quark and Lepton Physics (MOE) and Institute of Particle Physics, Central China Normal University, Wuhan 430079, China}

\author{Shengfeng Deng}
%\email[]{tauber@vt.edu}
\affiliation{Key Laboratory of Quark and Lepton Physics (MOE) and Institute of Particle Physics, Central China Normal University, Wuhan 430079, China}

\author{Tao Zhang}
\affiliation{Key Laboratory of Quark and Lepton Physics (MOE) and Institute of Particle Physics, Central China Normal University, Wuhan 430079, China}
\date{\today}
% It is always \today, today, but any date may be explicitly specified

\begin{abstract}

Machine learning (ML) has been well applied to studying equilibrium phase transition models, by accurately predicating critical thresholds and some critical exponents. Difficulty will be raised, however, for integrating ML into non-equilibrium phase transitions. The extra dimension in a given non-equilibrium system, namely time, can greatly slow down the procedure towards the steady state. In this paper we find that by using some simple techniques of ML, non-steady state configurations of directed percolation (DP) suffice to capture its essential critical behaviors in both (1+1) and (2+1) dimensions. With the supervised learning method, the framework of our binary classification neural networks can identify the phase transition threshold, as well as the spatial and temporal correlation exponents. The characteristic time $t_{c}$, specifying the transition from active phases to absorbing ones, is also a major product of the learning. Moreover, we employ the convolutional autoencoder, an unsupervised learning technique, to extract dimensionality reduction representations and cluster configurations of (1+1) bond DP.
It is quite appealing that such a method can yield a reasonable estimation of the critical point. 

% BrainPainter\footnote{Source code: \url{https://github.com/mrazvan22/brain-coloring}} is customisable, easy to use, and can run straight from the web browser: \url{http://brainpainter.csail.mit.edu}.
\end{abstract}
% diagram showing the aim: input numbers and output images
\maketitle

\section{Introduction}
\label{intro}
The ability of modern machine learning (ML) \cite{jordan2015machine}  technology to classify, identify, or interpret large number of data sets, such as images, presupposes that it is suitable for providing similar analysis involved in exponentially large data sets of complex systems in physics. Indeed recently ML techniques have been widely applied to various branches of physics, including statistical physics \cite{engel2001statistical,mehta2014exact}, condensed matter physics \cite{carleo2017solving,carrasquilla2017machine,wang2016discovering,van2017learning}, biophysics \cite{mckinney2006machine,schafer2014learning}, astrophysics \cite{vanderplas2012introduction}, and many more fields in physics \cite{mehta2019high,carleo2019machine}. ML techniques, especially the deep learning architecture \cite{goodfellow2016deep,krizhevsky2012imagenet} equipped with deep neural networks, have become very powerful and have shown great potential in discovering basic physics laws.

ML has been explored as an important tool for statistical physics by converting the lengthy Monte Carlo simulations into time-saving pattern recognition process.
In particular, ML has led to great success in the classification of phases and in identifying transition points of various protocol models such as the Ising model, potts model and XY model, etc\cite{carrasquilla2017machine,wang2016discovering,van2017learning,broecker2017machine,zhang2019machine}.  Generally, in statistical physics we use ML techniques in conjunction with configurations, of small-sized systems, generated by Monte Carlo simulations \cite{hammersley2013monte}. For instance, the supervised ML method has been implemented to study the Ising model \cite{carrasquilla2017machine} of equilibrium phase transitions, in which both the fully connected network (FCN) and convolutional neural network (CNN) are used to train and learn raw configurations of phases. The results indicated that ML could predict the critical temperature and the spatial correlation exponent of the Ising model.
In \cite{zhang2019machine}, the supervised ML method is also utilized to learn the percolation model and XY model with flying colors, and the critical threshold $p_{c}$ and the critical exponent $\nu$ are accurately predicted. These are just two typical examples of using the supervised ML method to study phase transitions. Of course, the study of unsupervised ML in phase transitions seems more appealing, despite its capability in lack of information  \cite{wang2016discovering,wetzel2017unsupervised,hu2017discovering,wang2017machine}.
Unsupervised ML techniques, like principal component analysis (PCA) and autoencoder, have been applied to dimensionality reduction representations of the original data and identifying distinct phases of matter. These unsupervised ML techniques are efficient in capturing latent information that can describe phase transitions.

So far the discussion is only concerned about the equilibrium phase transitions, most of which has been very well studied in regards of critical features and henceforth universality class.
Since the 1970's, non-equilibrium critical phenomena \cite{henkel2008non,hinrichsen2000non,lubeck2004universal,haken1975cooperative} have been attracting a lot of attention as they haven't been studied thoroughly yet. Unlike the equilibrium phase transitions, the universality class of non-equilibrium phase transitions remains mysterious. Neither do we know what universality classes we have, nor do we know how many of them there are. Although it is generally believed that there exists a large number of systems whose phase transition behaviors belong to the university class of directed percolation (DP), according to the so-called DP conjecture. Moreover, the means to handle the non-equilibrium phase transitions is very limited. Even for the simplest and most well-known non-equilibrium phase transition model, the DP process, its (1+1)-dimensional analytic solution is yet to be achieved \cite{hinrichsen2000non,lubeck2004universal}. Therefore, for low-dimensional systems (below the critical dimension $d_c$), the most reliable research method is still Monte Carlo simulations. However, the extra, namely time dimension extremely prolongs the procedure towards the steady state, which adds more difficulty in calculating the critical exponents and brings up more uncertainty in the measurements. Although some research methods for equilibrium critical phenomena \cite{amit2005field,domb1996critical} can also be applied to non-equilibrium phase transitions but they are mostly model specific. Apparently, compared to the equilibrium phase transitions, the non-equilibrium phase transitions are harder to be tackled. Henceforth the question is: can we find an efficient method to deal with non-equilibrium phase transitions,
in terms of the range of validity and the computation time? That is, the method should be both general and time saving.

In this work, we adopt both supervised learning and unsupervised learning to study the bond DP process. For the supervised learning techniques, the neural networks we use here are FCN and CNN \cite{goodfellow2016deep}, which are fit for 2-D image recognition. One of the findings is that the non-steady state configurations are sufficient for capturing the essential information of the critical states, in both (1+1)- and (2+1)-dimensional DP. Naturally, our neural networks can train and detect some latent parameters and exponents by sampling the configurations generated by Monte Carlo simulations \cite{Dhar_1981,grassberger1989directed} of small-sized systems. For the unsupervised learning, we employ the convolutional autoencoder to extract dimensionality reduction representations and cluster configurations of (1+1) bond DP.

The main structure of this paper is as follows. In Sec.\uppercase\expandafter{\romannumeral2}.A, we briefly introduce the model of bond DP. Sec.\uppercase\expandafter{\romannumeral2}.B displays the basic architectures of two neural networks that will be used in the supervised learning. Sec.\uppercase\expandafter{\romannumeral2}.C is the data sets and ML results of (1+1)- and (2+1)-dimensional bond DP. Here, we successfully confirmed the critical points, the temporal and spatial correlation exponents, and the characteristic time exponent. In Sec.\uppercase\expandafter{\romannumeral3}, the autoencoder method of unsupervised ML is implemented to extract dimensionality reduction representations and cluster the generated configurations of (1+1)-dimensional bond DP, from which we can estimate the critical point. Finally, Sec.\uppercase\expandafter{\romannumeral4} is a summary of this paper.

\section{Supervised Learning of Directed Percolation}
\subsection{The Model of DP}
The non-trivial DP lattice model is simple and easy to simulate. The phenomenological scaling theory can be applied to describing DP, by which critical exponents are determined. Once the whole set of critical exponents of a model is complete, the universality class that the model belongs to establishes.

The DP model, exhibiting a continuous phase transition, is the most prominent universality class of absorbing phase transitions \cite{grinstein1997statistical}, which can describe a wide range of non-equilibrium critical phenomena such as contact processes \cite{liggett2012interacting,dickman1988nonequilibrium}, forest fires \cite{clar1996forest,albano1994critical}, epidemic spreadings \cite{mollison1977spatial}, as well as catalytic reactions \cite{ziff1986kinetic}, etc. The conventional approaches for studying the scaling behaviors of DP are mean-field theory \cite{de1983directed}, renormalization groups \cite{kinzel1981directed}, field theory approach \cite{janssen2005field} and numerical simulations \cite{hinrichsen1998numerical} and so on. Generally, for DP, the mean-field approximation is valid above the upper critical dimension $d_{c} = 4$, as the particles diffusion mix exactly well. In contrast, the field theory method is applicable below the upper critical dimension. For low-dimension cases, numerical simulations are usually more convenient to implement. In the numerical simulation of DP, to obtain relatively accurate results, we usually take a large system size (for instance, $L \geq 10000$). This condition, in turn, causes the characteristic time $t_{c} \sim L^{z}$ \cite{hinrichsen2000non} of the system being too large such that it takes extremely long to approach the steady states. This is the tricky thing that one has to deal with when simulating non-equilibrium systems.
Given the advantages of ML techniques over small-sized systems, this paper will employ them to learn the bond DP model. For small-sized systems, in particular, it turns out that ML can quickly and accurately determine critical thresholds and critical exponents of many phase transition models.

\begin{figure*}[t]
\centering

$\xrightarrow[]{position \quad  i}$

\vspace{1.6ex}

\rotatebox{90}{$\xleftarrow[]{time \quad  t}$}
\hspace{0.8ex}
\includegraphics[scale=0.4]{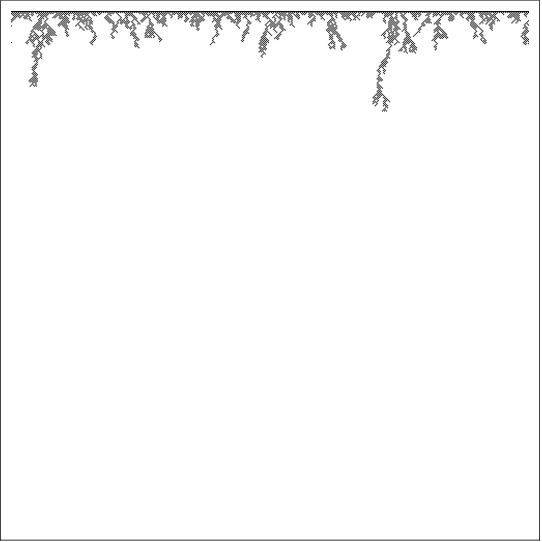}
\includegraphics[scale=0.4]{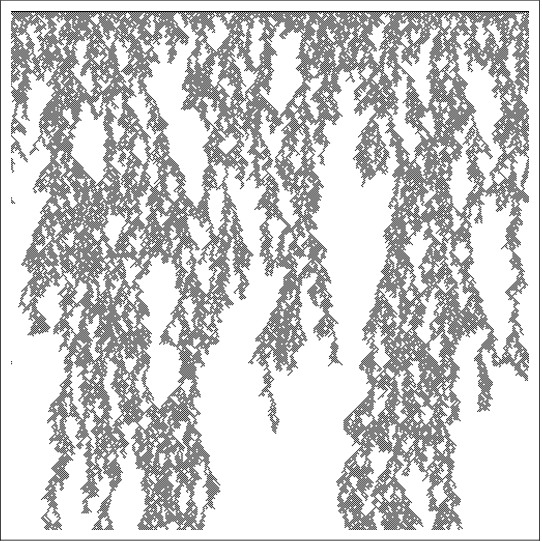}
\includegraphics[scale=0.4]{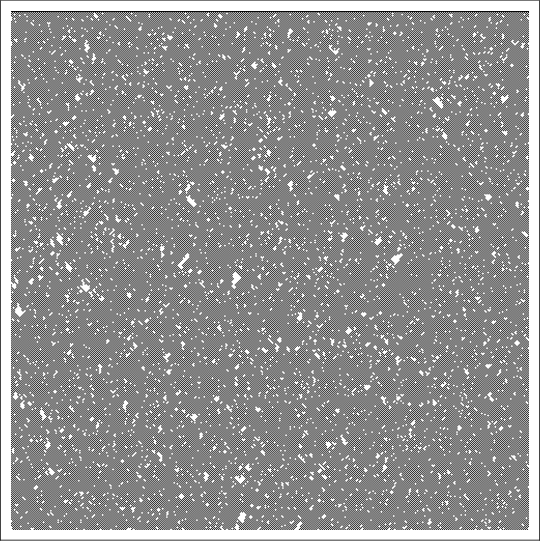}

\hspace{4.2ex}
\includegraphics[scale=0.4]{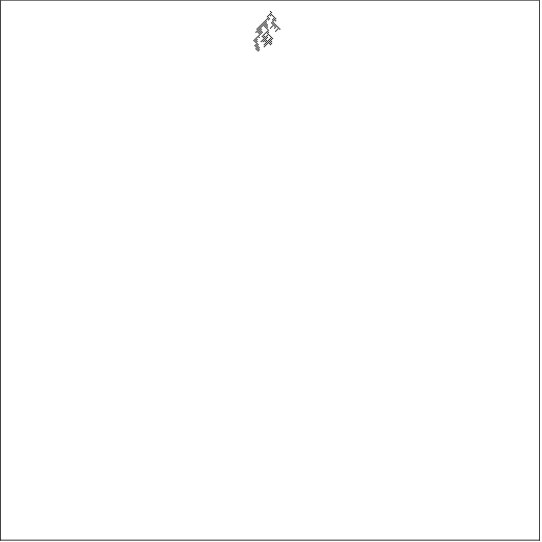}
\includegraphics[scale=0.4]{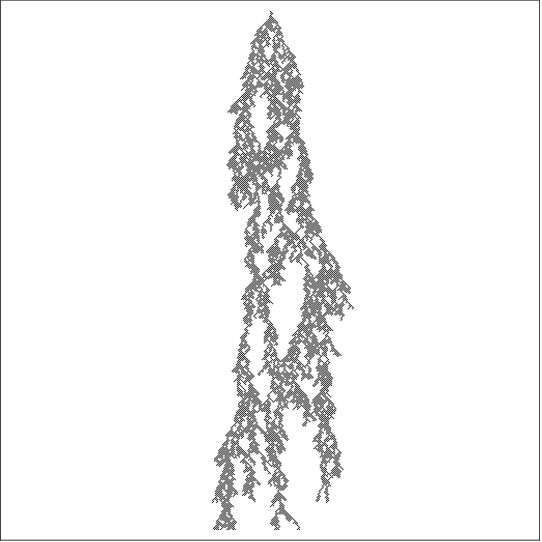}
\includegraphics[scale=0.4]{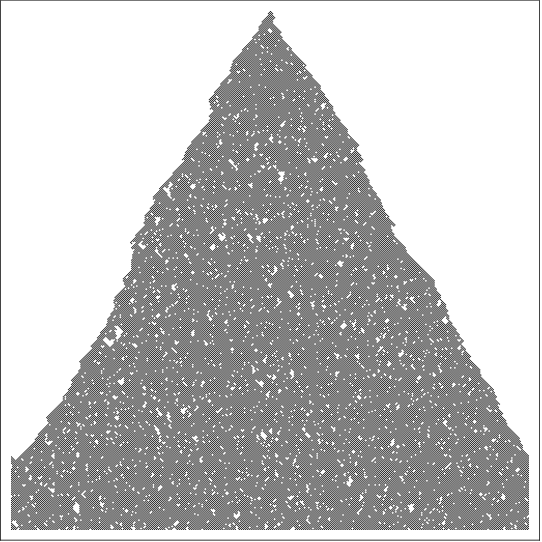}
\caption{Directed bond percolation in (1+1) dimension, starting from a fully occupied lattice (top panel) and from a single active seed (bottom panel), where $L = 500$, and the bond probability $p$ is $0.55$, $0.6447$, and $0.8$, respectively.}
\label{fig:single}
\end{figure*}

The main work of this paper is to apply ML techniques to (1+1)- and (2+1)- dimensional bond DP \cite{grassberger1989directed,maslov1996self,wang2013high} of non-equilibrium phase transitions. Before implementing ML, it is necessary to introduce the bond DP model. The model goes in two ways, either starting from a fully occupied initial state or from a single active seed (configurations are shown in Fig. \ref{fig:single}). Then the model evolves through intermediate configurations according to the dynamic rules of the following equations,
\begin{equation}
  s_{i}(t+1)=\left\{
\begin{array}{rcl}
1     &      & {if \quad s_{i-1}(t) = 1 \quad and \quad z_{i}^{- } < p,}\\
1     &      & {if \quad s_{i+1}(t) = 1 \quad and \quad z_{i}^{+ } < p,}\\
0     &      & {otherwise,}
\end{array} \right.
\end{equation}
where $s_{i}(t+1)$ represents the state of node $i$ at time  $t + 1$, $z_{i}^{\pm} \in (0, 1) $ are two random numbers. Configurations of two types of (1+1)-dimensional bond DP are shown in Fig. \ref{fig:single}, where two adjacent sites are connected by a bond with probability $p$. If the bond probability is large enough, a connected cluster percolates throughout the system.

The evolution of DP configurations is closely related to bond probability. In the case of a fully occupied lattice, for $p < p_{c}$, the number of active particles will decay exponentially, and eventually the system reaches an absorbing phase. On the contrary, for $p > p_{c}$, the number of active particles will reach a saturation value and we say the system is in an active phase. At the critical point of $p = p_{c}$, the number of particles decreases slowly, following a power-law. On the other hand, how the bond DP starting from a single active seed evolves is shown in the bottom panel of Fig. \ref{fig:single}. For $p < p_{c}$, the number of active particles increases for a short time and then decays exponentially, while for $p > p_{c}$ the probability to form an infinite cluster is finite. At the critical point $p = p_{c}$, an infinite cluster will be formed.

The critical behavior of DP can be described in two different ways, either by percolation probability, or via the order parameter of steady-state density,
\begin{equation}
  P_{perc}(p)\widetilde{\propto}(p-p_{c})^{\beta^{^{\prime}}},\quad
 \rho_{a}(p)\widetilde{\propto}(p-p_{c})^{\beta}
\end{equation}
where $P_{perc}(p)$ represents the probability that a site belongs to a percolating cluster, $\rho_{a}$ denotes particle density, and $\beta=\beta^{\prime}$ is resulted by rapidity-reversal symmetry. There are some other observations in the DP model, such as the number of active sites $N(t,p)$, the survival probability $P(t,p)$, and the average mean square distance of the spreading $R^{2}(t,p)$. Besides, we also attach importance to two non-independent correlation lengths, which are called temporal correlation length $\xi_{\parallel}$ and spatial correlation length $\xi_{\perp}$. At criticality, they exhibit the following asymptotical power-law behaviors
  \begin{equation}
   N(t,p_{c})\sim t^{\eta},   \quad P(t,p_{c})\sim t^{-\delta}, \quad R^{2}(t,p_{c})\sim t^{2/z},
  \end{equation}

 \begin{equation}
  \xi_{\parallel}\sim \mid p-p_{c}\vert^{-\nu_{\parallel}} ,   \quad  \xi_{\perp}\sim \mid p-p_{c}\vert^{-\nu_{\perp}}.
  \end{equation}
where $\nu_{\parallel}$ and $\nu_{\perp}$ are the temporal correlation exponent and the spatial correlation exponent, respectively.

In non-equilibrium statistical mechanics, systems evolve over \textit{time} which is an independent degree of freedom on equal footing with the spatial degrees of freedom. When a fully occupied lattice as an initial state is specified for a system, close to the critical state, the density of active sites decays algebraically as
\begin{equation}
\varrho(t) \sim t^{-\delta},
\label{delta}
\end{equation}
where $\delta = \beta / \nu_{\parallel}$. Similarly, the spatial correlation length grows as
\begin{equation}
\xi_{\perp}(t) \sim t^{1/z},
\end{equation}
where $z = \nu_{\parallel} / \nu_{\perp}$. In finite-size systems, one finds deviations from these asymptotic power-laws. There will be finite-size effects after a typical time which grows with the system size. If $L$ is the lateral size of the system (and $N \varpropto L^{d}$ is the total number of sites), the absorbing state reaches at a characteristic time $t_{c} \sim L^{z}$ \cite{henkel2008non}.

The critical exponents of DP are not independent and satisfy the following scaling relations,
\begin{equation}
   \delta=\beta/\nu_{\parallel},   \quad z=\nu_{\parallel}/\nu_{\perp},\quad \eta=\dfrac{d}{2}-2\dfrac{\beta}{\nu_{\parallel}}.
\label{zzzzz}
\end{equation}
Unlike isotropic percolation, the upper critical dimension of DP is $d_{c} = 4$ \cite{obukhov1980problem,cardy1980directed}. The scaling relation with the dimension $d$ holds only if $d \leq d_{c}$. The critical thresholds and scaling exponents of DP can be checked in Table. \ref{table2}.

\subsection{Methods of Machine Learning}
\begin{figure*}[t]
    \centering
    \begin{minipage}{0.6\textwidth}
        \centering
        \includegraphics[width=1.1\textwidth]{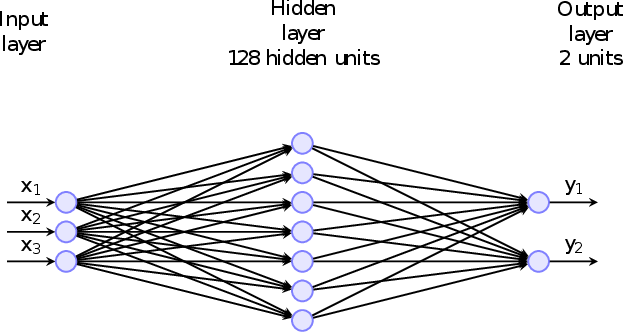} % first figure itself
        \centerline{a}
%        \caption{first figure}
    \end{minipage}\vfill
    \begin{minipage}{0.7\textwidth}
        \centering
        \includegraphics[width=1.2\textwidth]{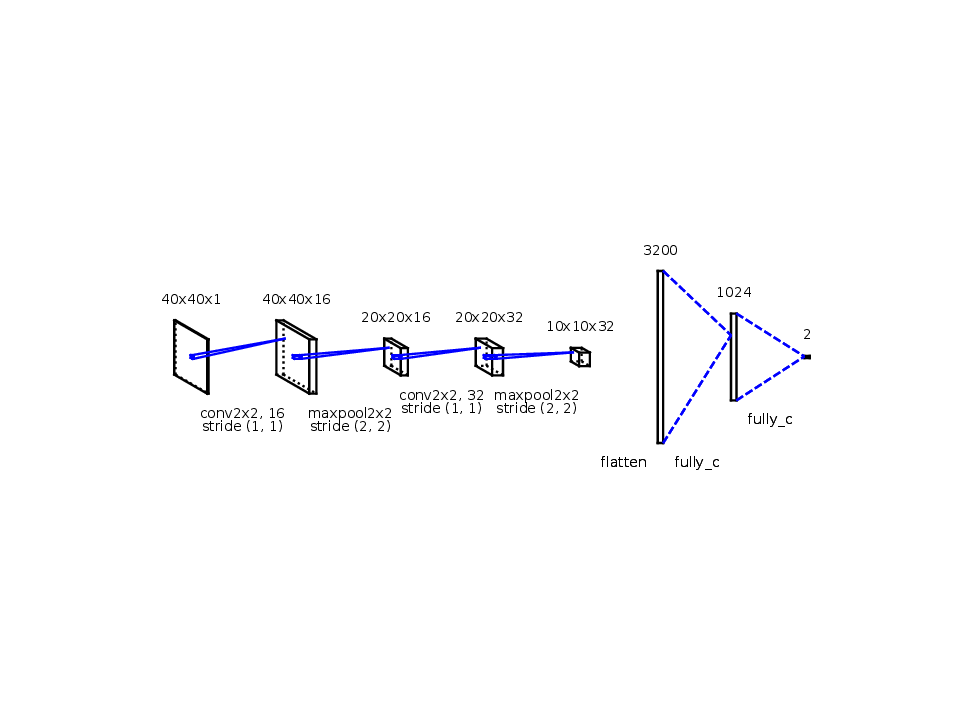} % second figure itself
        \centerline{b}
%        \caption{second figure}
    \end{minipage}
\caption{Schematic structure of two neural networks. Panel a displays a fully connected network (FCN) and panel b, a two-dimensional convolutional neural network (CNN).}
\label{fcncnn}
\end{figure*}

Usually, one uses ML methods, which include supervised learning, unsupervised learning, reinforcement learning, and so on, to recognize hidden patterns embedded in complex data. With the continuous improvement of computing power, neural network based algorithms play significant roles in ML. Extremely complex tasks usually require deep neural networks, which contain many hidden layers. The two main tasks of supervised learning are regression and classification. Before classification learning, we implemented a linear regression machine learning on DP's critical configurations. To classify the phases of DP, all we need is a simple neural network. The neural networks we use here are the most common and widely used, FCN and CNN, as shown in Fig. \ref{fcncnn}. These two networks have excellent learning effects for image recognition and classification. In our neural networks, the input data is the bond DP configurations generated from Monte-Carlo simulations of small-sized systems.

In artificial neural networks, the structure of FCN generally consists of an input layer, a hidden layer, and an output layer. Its neurons are fully connected to the others of the previous layer and the next layer. For our FCN, the hidden layer contains $128$ neurons, and two neurons are in the output layer so that we can carry out a binary classification between active phases and absorbing phases.

Our CNN structure consists of two layers, convolutional layer and pooling layer. Actually we do not have to limit ourselves to two layers, and we could instead employ a stack of convolutional and pooling layers if needed. The images that are fed into CNN are DP configurations. For each of them, $LENGTH$ of an image represents the lattice length of DP, and $TIME$ is the time step. The convolution kernel for convolving the two-dimensional images is also two dimensional. The padding form in CNN takes the form of "same", which keeps the size of the feature map of the image after the convolution operation. As a two-dimensional convolution kernel can only extract one feature in the convolution operation of the whole two-dimensional image, henceforth the kernels share weights. So we can use multiple convolution kernels to extract multiple features. After the feature map processing through the convolutional layer and pooling layer, the data flow is flattened and then transmitted to a fully connected layer. Finally, we take a binary classification via the fully connected layer to produce the classification result. To prevent over-fitting \cite{glassner2018deep}, we add the L2-norm ($\lambda/(2N) \sum\limits_{i} w^{2}_{i}$) to the loss function. The AdamOptimizer is used to speed up our neural networks. Our ML is implemented based on TensorFlow 1.15.

\subsection{Data Sets and Results}
In supervised learning, the data sets generated by Monte Carlo simulations include a training set, a validation set, a test set and their corresponding label sets. Among them, the generating mechanisms of the training, validation and test sets are the same. And the use of the validation set is to optimize our model parameters.

Before learning, the generated data of configurations need to be labeled, and some hyper-parameters of neural networks are defined. In the input layer, the number of neurons equals the dimensions of the input data, which is $L \times (t + 1)$ ($t$ represents the percolation time step). An input configuration generated with the bond probability $p < p_{c}$ is labeled as "0", which means that the system reaches an absorbing phase. For $p > p_{c}$, an input configuration is labeled as "1", and the system is thought of as percolating. Here, five different system sizes of the fully occupied initial state are used to generate data, where $L$ is 8, 16, 32, 48, and 64, respectively. For each size, in (1+1)-dimensional DP, the configuration is generated by selecting $41$ bond probabilities between $(0.4,0.9)$ with an interval of $0.0125$, and $2500$ samples per probability. The ratio of training set, verification set and test set samples is $5:2:2$. The bond probability range in the (2+1)-dimensional case is $(0.1,0.5)$, where we also use a similar sampling mechanism.

In order to study the effect of input data transformation on supervised ML, we will use FCN and CNN respectively to learn the original data after transformation. The data transformation for DP is as follows. In the case of (1+1)-dimensional DP, we flatten the data of each time step into a one-dimensional array and then feed it to FCN. And CNN, on the other hand, is fed a two-dimensional array. In the case of (2+1) dimensions, the configuration data of DP is also flattened into a one-dimensional array and then fed to FCN. The data that CNN receives is still a two-dimensional array transformed from a three-dimensional image. In addition, in the case of (1+1) dimensions, a configuration is just an image or a two-dimensional array, and we do not need to remove or filter any input data, but intercept a part of it. Through testing, we find that even if we only select configurations corresponding to a part of the time series, we are able to learn the correct results. Namely, the characteristic time of DP is $t \sim L^{z}$, which signifies the zone of steady-state, but we only choose $l_{x} * l_{y} $ ($l_{x} = L $, $l_{y}  = L = t_{real} + 1 $) to learn, where $l_ {x} $ and $l_ {y} $ respectively represent the length and width of the input data. This choice greatly reduces the computational effort.

%Among them, more dense bond probabilities are selected near the critical point $p_{c} = 0.6447$ \cite{hinrichsen2000non}, for the sake of a more accurate training effect.

Fig. \ref{fig:single} shows two configurations generated from two different initial conditions under the same system size. It is natural to think which one will be more applicable for the learning. After systematic tests, we found that configurations generated by a fully occupied initial state are more suitable for our algorithm. This is empirically evidenced by the testing accuracy of the initial condition with a single active seed as is also presented in Tab. \ref{table:15}. It can be clearly seen that the fully occupied initial state has higher testing accuracy than the single-seed one. We conjecture that the fully occupied initial state possesses more sites, which can provide much more information to the neural network so that the ML results obtained are more accurate. Therefore, the following discussions are based on configurations generated from a fully occupied initial state.

\subsubsection{Learning DP's critical configuration via linear regression}

\begin{figure*}[t]
%\begin{tabular}{cc}
\begin{tabular}{cc}
    \includegraphics[width=0.45\textwidth]{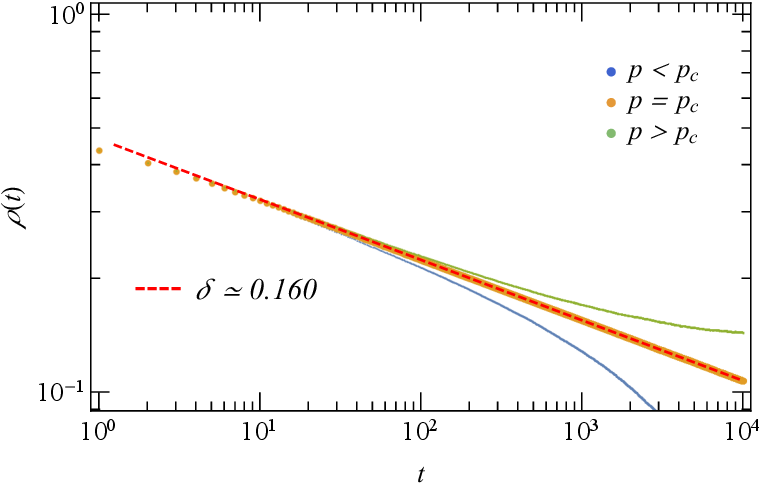} &
    $\qquad$\includegraphics[width=0.50\textwidth]{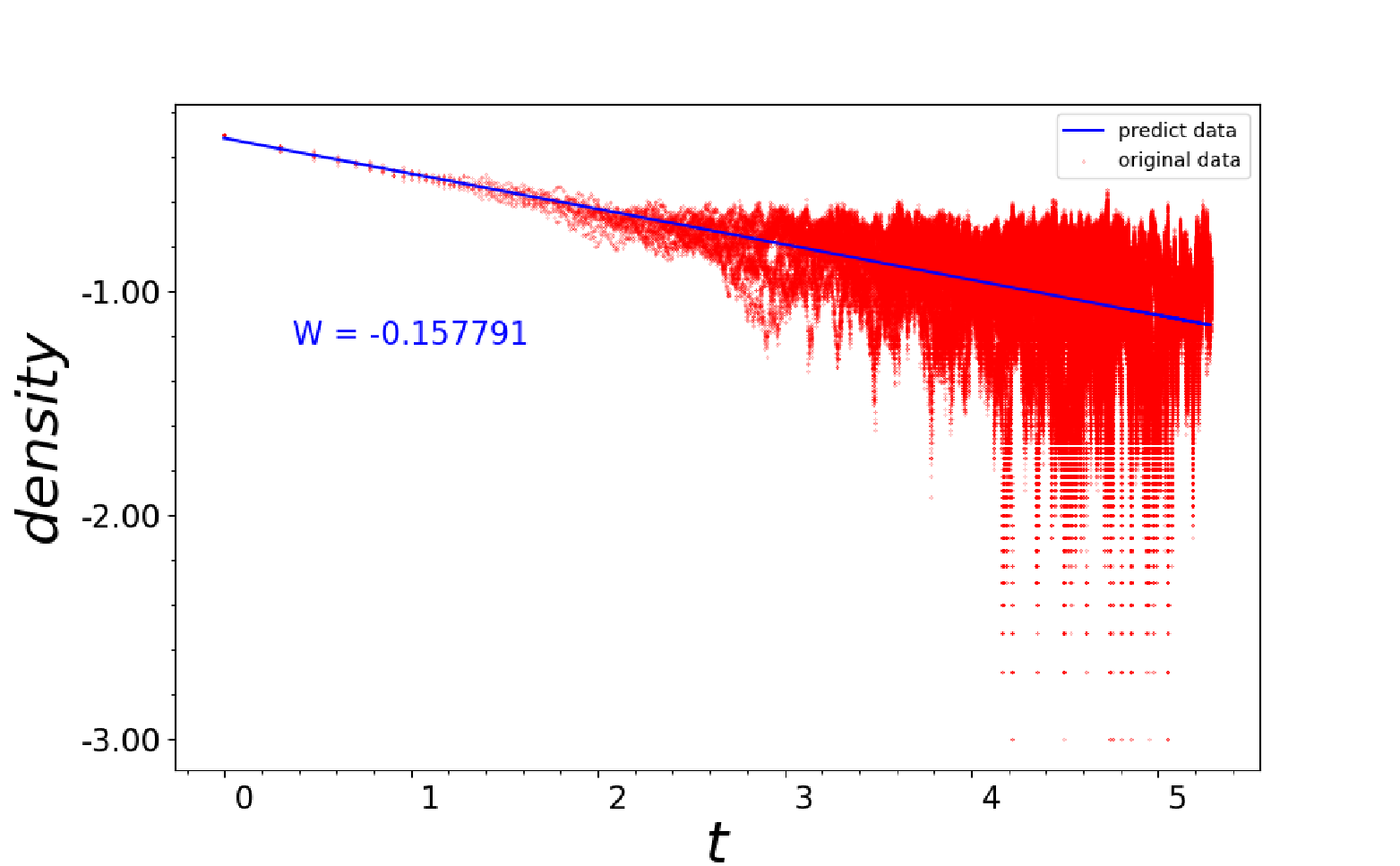} \\
    (a) & $\qquad$ (b)
\end{tabular}
\caption{ \textbf{a}, (1+1)-dimensional DP Monte Carlo simulations, where the lattice size is $L=12000$, the total time step is $t=10000$, and the number of ensemble average is $10000$, for different $p=0.6425(blue), 0.6447(yellow), 0.6460(green)$, respectively. The red dashed line is a fitting with $p_{c}=0.6447$, whose slope is $\delta =0.160$. \textbf{b}, The linear regression result of (1+1)-dimensional DP, where the total time step is $t= L^{1.58}$, the lattice size is $L=1000$, and the number of runs is $10$.}
\label{lire}
\end{figure*}

If one needs to understand the meaning of the time variable in DP, the particle density at a time step, namely the ratio of the occupied lattice divided to the total number of lattice points, could be measured.

The left panel of Fig. \ref{lire} shows the Monte Carlo simulation results of (1+1)-dimensional bond DP with a fully occupied initial state. According to Eq. \ref{delta}, the critical exponent $\delta \simeq 0.160$ can be obtained by fitting the particle density $\rho(t)$ versus $t$. With this as a reference, we now perform a linear regression for the critical configuration of DP, as shown in the right panel of Fig. \ref{lire}. Our original purpose is to predict $\rho(t)$ at a certain moment by the linear regression method of machine learning. That is to say, fed an input of $\rho(t)$ at an early time step, the regression can predict $\rho(t)$ at a later time step. In (b) of Fig. \ref{lire}, after $20000$ iterations, the weight learned from the regression model is $W = -0.157791$, that is, $\delta = 0.157791$. As can be seen, a single time step can correspond to many different particle densities, so the exponent $\delta$ obtained by such a regression model is not accurate. The main reason is that the critical configuration fluctuates greatly. Therefore, for DP, we believe that linear regression is not a good tool.

Next, we move to other critical exponents of DP.

\subsubsection{Learning DP phases via FCN}
The main goal of using supervised ML in this article is to make phase classification of DP configurations generated by different bond probabilities, from which the critical properties, such as critical points and critical exponents, can be obtained. This target is very different from time series classification or prediction, for instance the studies given in Refs. \cite{2001Time,2018Time}. Although our model produces time series, we are dealing with phase classification rather than time series classification. Namely, all the configurations generated by a single bond percolation probability belongs to the same category. For example, in the case of (1+1)-d DP, the configuration corresponding to a certain time step is one-dimensional. We combine the configurations of the all-time series into a two-dimensional image for machine learning. Therefore, in our study time is just another dimension.

\begin{figure*}[t]
%\begin{tabular}{cc}
\begin{minipage}{0.3\linewidth}
  \centerline{\includegraphics[width=6.0cm]{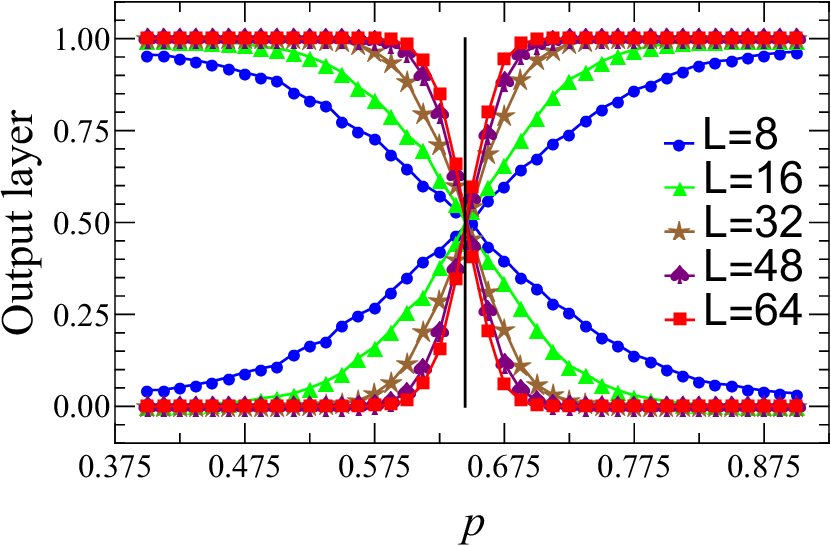}}
  \centerline{a}
\end{minipage}
\hfill
\begin{minipage}{0.3\linewidth}
  \centerline{\includegraphics[width=6.0cm]{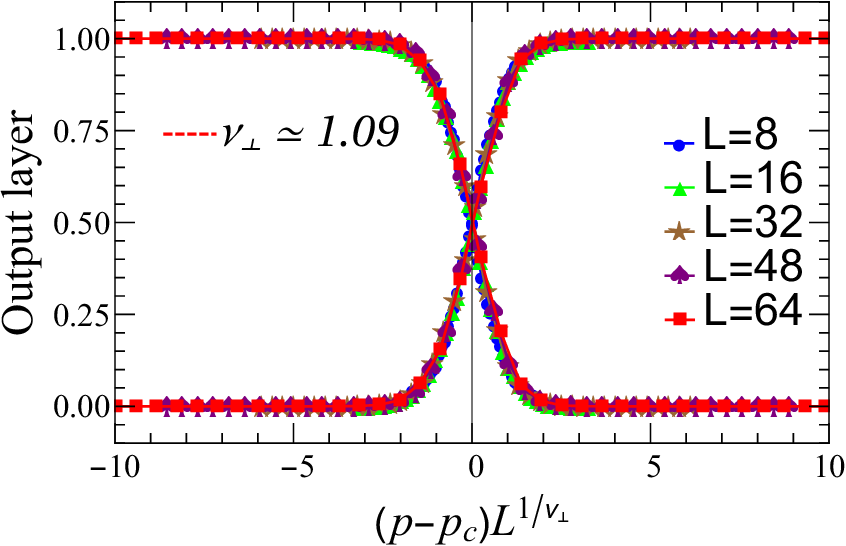}}
  \centerline{b}
\end{minipage}
\hfill
\begin{minipage}{0.3\linewidth}
  \centerline{\includegraphics[width=6.0cm]{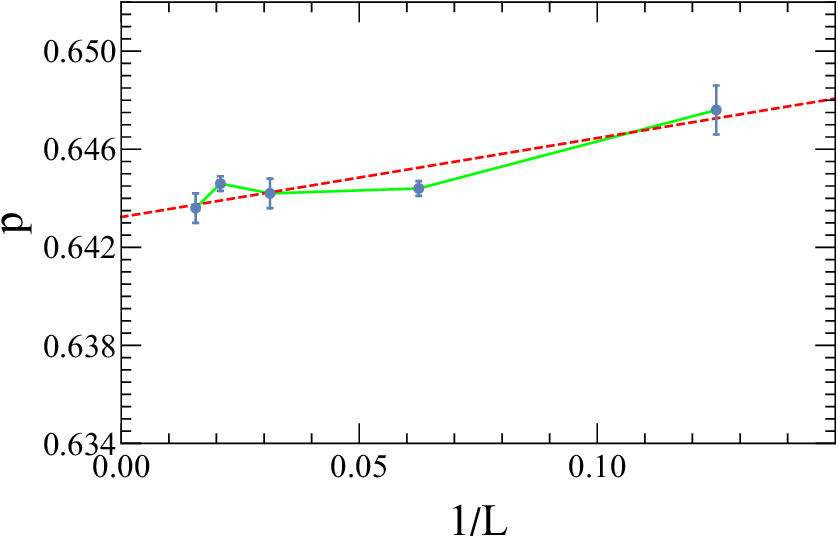}}
  \centerline{c}
\end{minipage}
\vfill
\begin{minipage}{0.3\linewidth}
  \centerline{\includegraphics[width=6.0cm]{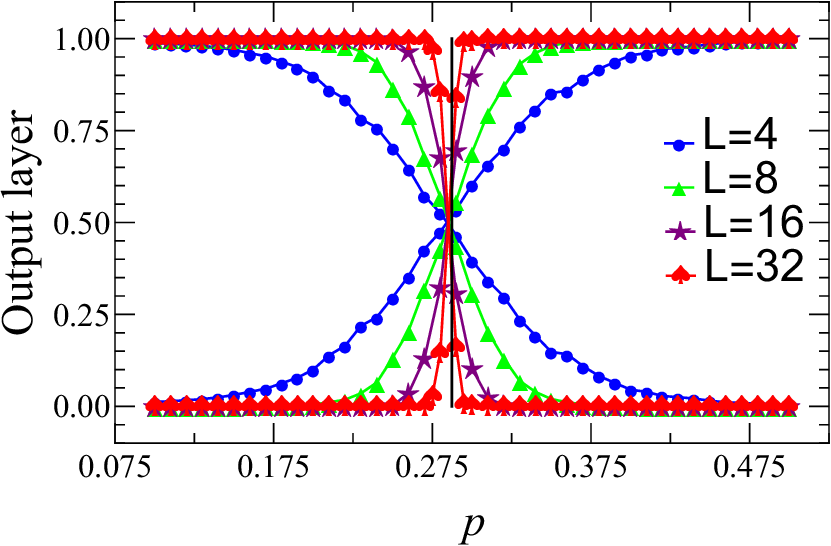}}
  \centerline{d}
\end{minipage}
\hfill
\begin{minipage}{0.3\linewidth}
  \centerline{\includegraphics[width=6.0cm]{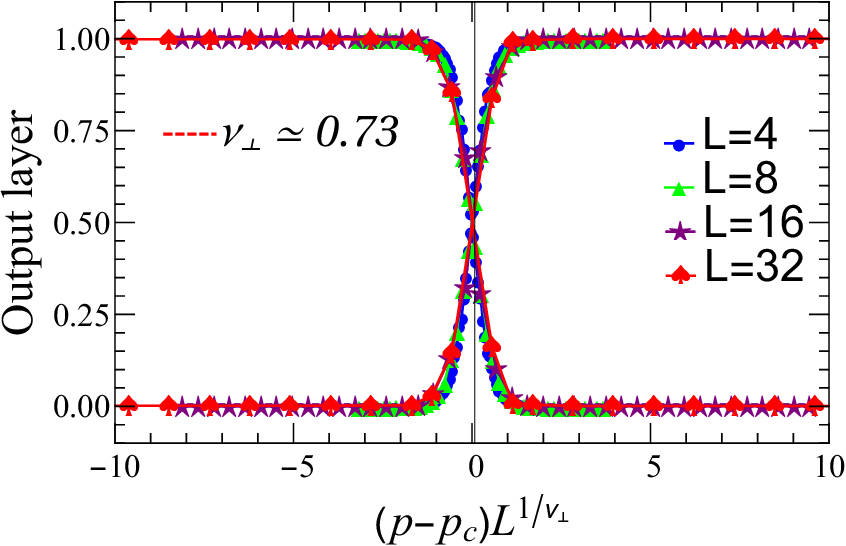}}
  \centerline{e}
\end{minipage}
\hfill
\begin{minipage}{0.3\linewidth}
  \centerline{\includegraphics[width=6.0cm]{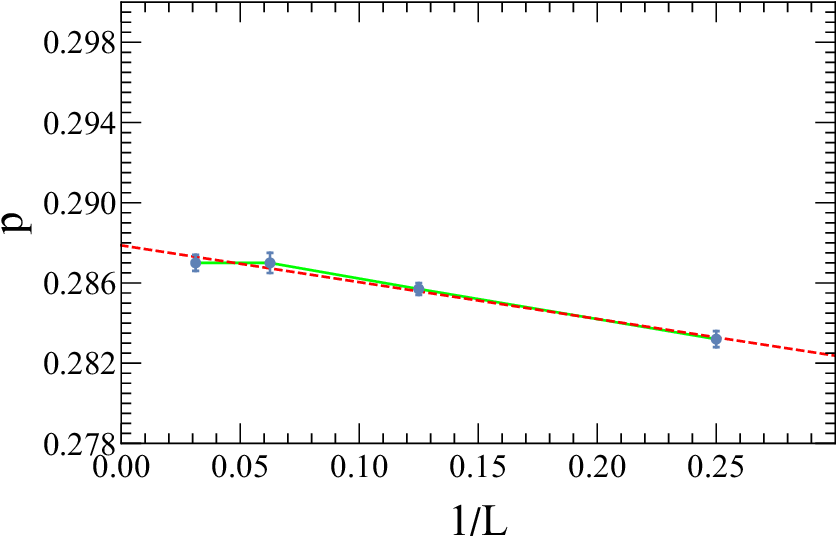}}
  \centerline{f}
\end{minipage}
%\end{tabular}
\caption{ML (1+1)-dimensional (top panel) and (2+1)-dimensional (bottom panel) bond DP by FCN. \textbf{a}, The output layer, averaged over a test set, as a function of the bond probability $p$. \textbf{b}, Data collapse of the average output layer as a function of $( p- p_{c}) L^{1/\nu_{\perp}}$. System sizes of $L = 4$, 8, 16, 32, 48, and 64 are represented by different colors, respectively. \textbf{c}, Plot of the finite-size scaling of the bond probability $p_{c}$. \textbf{d}-\textbf{f}, Analogous data to \textbf{a}-\textbf{c}. The vertical black lines signal the critical thresholds of two cases, $p_{c} = 0.6447$ for the (1+1) dimensions and $p_{c} = 0.2873$ for the (2+1) dimensions. The error bars manifest statistical uncertainty of standard deviation of the mean.}
\label{fig:res}
\end{figure*}

Fig. \ref{fig:res} shows the ML results of (1+1)-dimensional and (2+1)-dimensional bond DP by using FCN. In panel (a) of Fig. \ref{fig:res}, the mean prediction of the output layer is given for (1+1)-dimensional DP, which distinguishes DP configurations in the range of $0.4  < p < 0.9$ for five different system sizes, with $L$ = 8, 16, 32, 48, and 64. We illustrate with a pair of blue curves for $L=8$, which correspond to the two output predictions of the neural network. As $p$, the bond probability, increases, the prediction probability that the configurations generated by the corresponding bond probability are in an active phase also increases. Whereas, the other curve indicates that with the increasing of $p$, the prediction probability that the configuration will reach an absorbing phase decreases continuously. And the intersection point of these two curves is the phase transition point predicted by the neural network \cite{carrasquilla2017machine,zhang2019machine}. Since all configurations are generated by finite-size lattices, the range of $L$ will restrict the length of connectedness, and no real phase transition occurs with limited degrees of freedom. So for the spatial correlation length, near the phase transition point we have,
\begin{equation}
\xi_{\perp} \sim L \sim \left| p- p_{c} \right|^{-\nu_{\perp}} \rightarrow    \left| p- p_{c} \right| \sim L^{-1/\nu_{\perp}}.
\end{equation}

In order to eliminate the influence of finite-size effects, the abscissa $p$ of Fig. \ref{fig:res} (a) is rescaled to $(p-p_{c})L^{1/\nu_{\bot}}$, which yields Fig. \ref{fig:res} (b). By tuning $\nu_{\perp}$ to make all five pairs of curves collapse, we obtain $\nu_{\perp} \simeq 1.09 \pm 0.02$. We notice that in Fig. \ref{fig:res} (b) the blue curves appear a little bit bumped compared to other curves, due to its small lattice size. Then, taking out the five crossing critical thresholds predicted in five different system sizes, we plot them as a function of $1/L$ to achieve finite-size scaling with extrapolation from $L$ being finite to $L \sim \infty$. As shown in Fig \ref{fig:res} (c), the critical point is $p_{c} \simeq 0.6433$ through finite-size scaling fitting.

By using the same output layer of the above five systems, one can also get the temporal correlation exponent in the (1+1)-dimensional bond DP, provided that the dynamic exponent $z$, in Eq. \ref{zzzzz}, is known. The scaling governing the temporal correlation length is given by,
\begin{equation}
\xi_{\parallel} \sim L^{z} \sim \left| p- p_{c} \right|^{-\nu_{\parallel}} \rightarrow \left| p- p_{c} \right| \sim (L^{z})^{-1/\nu_{\parallel}},
\end{equation}
where the temporal correlation exponent $\nu_{\parallel} \simeq 1.73 \pm 0.02$ can be obtained by the data collapse involving rescaling the abscissa into $(p- p_{c})(L^{z})^{1/\nu_{\parallel}}$.

The critical threshold of (1+1)-dimensional bond DP, through learning via FCN, conforms to the counterpart given by the literature \cite{hinrichsen2000non}. Furthermore, the spatial and temporal correlation exponents of DP can be obtained by the data collapse of finite-size scalings.

In addition, the FCN is also applied to learn and predict the (2+1)-dimensional bond DP, and the results are shown in the bottom row of Fig. \ref{fig:res}. For the sake of calculation, the time steps are still selected to be $t_{real} = L$ ($t \sim L^{z}$). It's worth noting that in (1+1) dimensions we learn and predict the configurations of dimension $l_{x} \times l_{y}$, where $l_{x} = L$, $l_{y} = t_{real} + 1$. But for the case of (2+1) dimensions, we need to transformed its configurations from $L \times L \times T$ into $l_{x} \times T$. The fact is that, we reshape $l_{x} = L \times L$ and $l_{y} = t_{real} + 1$ into a two-dimensional image as the input configuration. As expected, the results of the (2+1)-dimensional bond DP indicates that although the input data are reshaped, the FCN can still learn the trend of configurations.

\subsubsection{Learning DP phases via CNN}
\begin{figure*}[!tbp]
%\begin{tabular}{cc}
\begin{minipage}{0.3\linewidth}
  \centerline{\includegraphics[width=6.0cm]{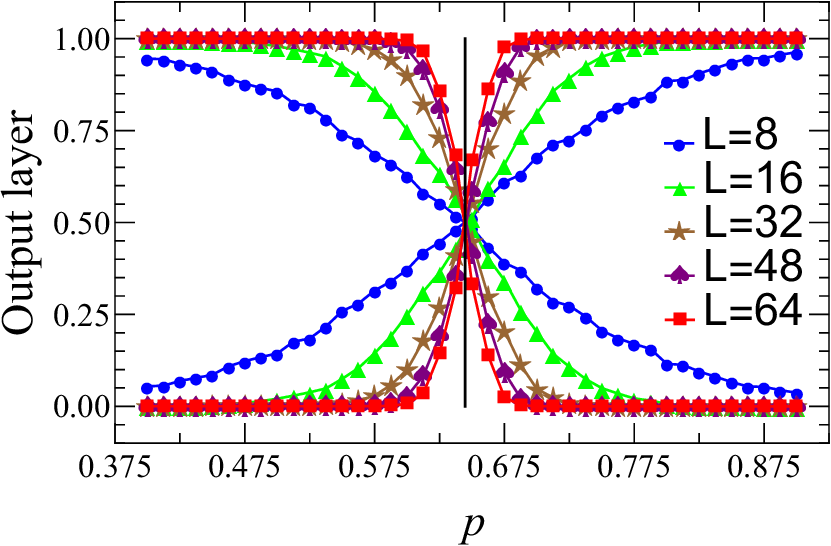}}
  \centerline{a}
\end{minipage}
\hfill
\begin{minipage}{0.3\linewidth}
  \centerline{\includegraphics[width=6.0cm]{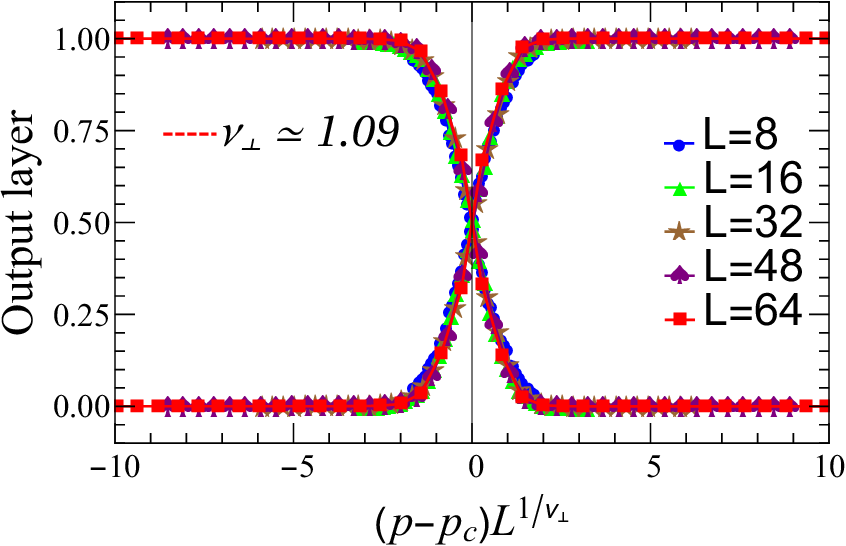}}
  \centerline{b}
\end{minipage}
\hfill
\begin{minipage}{0.3\linewidth}
  \centerline{\includegraphics[width=6.0cm]{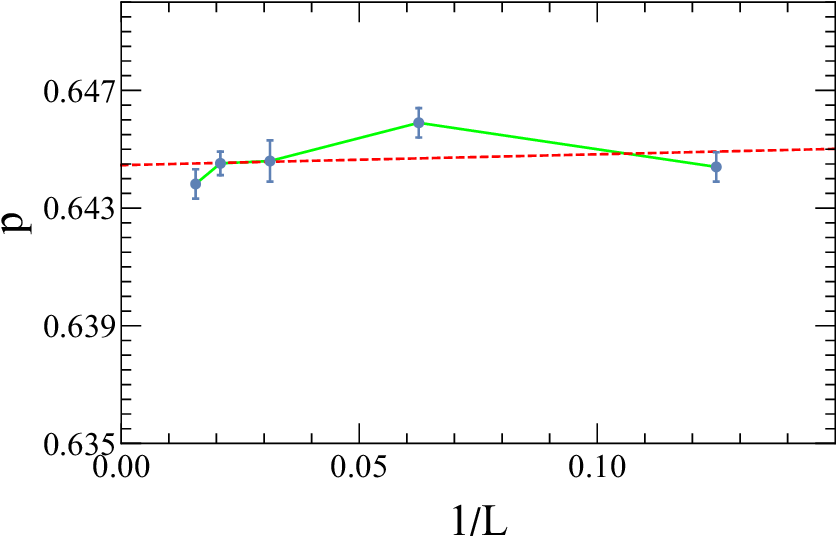}}
  \centerline{c}
\end{minipage}
\vfill
\begin{minipage}{0.3\linewidth}
  \centerline{\includegraphics[width=6.0cm]{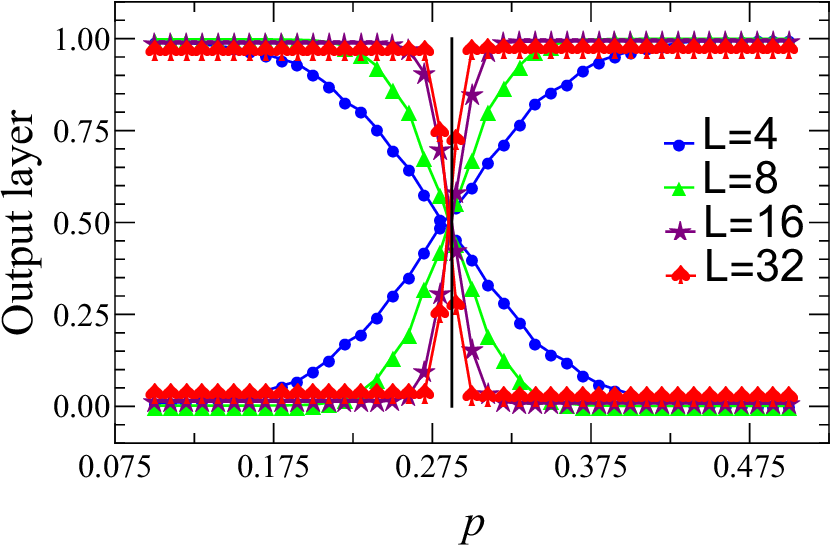}}
  \centerline{d}
\end{minipage}
\hfill
\begin{minipage}{0.3\linewidth}
  \centerline{\includegraphics[width=6.0cm]{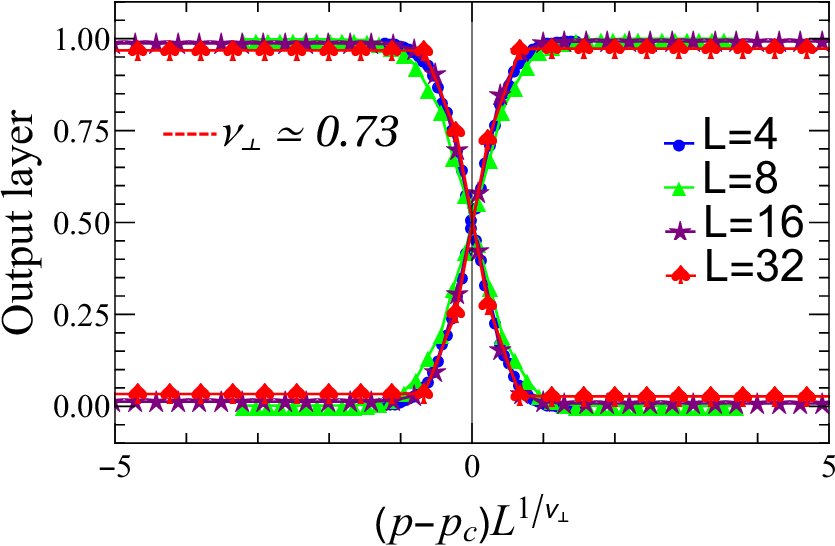}}
  \centerline{e}
\end{minipage}
\hfill
\begin{minipage}{0.3\linewidth}
  \centerline{\includegraphics[width=6.0cm]{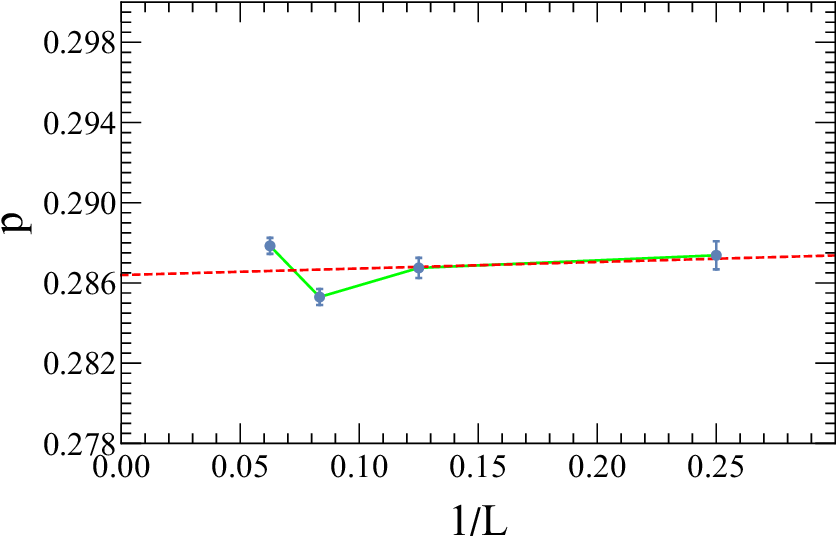}}
  \centerline{f}
\end{minipage}
%\end{tabular}
\caption{ML (1+1)-dimensional (top) and (2+1)-dimensional (bottom) bond DP by CNN. \textbf{a}, The output layer, averaged over a test set, as a function of the bond probability $p$. \textbf{b}, Data collapse of the average output layer as a function of $( p- p_{c}) L^{1/\nu_{\perp}}$. System sizes $L = 4, 8, 12, 16, 32, 48, 64$ are represented by different colors, respectively. \textbf{c}, Plot of the finite-size scaling of the bond probability $p_{c}$. \textbf{d}-\textbf{f}, Analogous data to \textbf{a}-\textbf{c}. The vertical black lines signal the critical thresholds of two models, $p_{c} = 0.6447$ for the (1+1) dimensions and $p_{c} = 0.2873$ for the (2+1) dimensions. The error bars represent statistical uncertainty of standard deviation of the mean.}
\label{mlcnn}
\end{figure*}

In this part, we present the learning results of bond DP by using CNN, as shown in Fig. \ref{mlcnn}. For the configurations of (1+1)-dimensional DP, we convert the one-dimensional lattice of each time step into a two-dimensional image. And for the case of (2+1)-dimensional DP, we reshape $L \times L \times T$ into $l_{x} \times T$ ($l_{x} = L \times L$). It turns out that the reshaped data still contain essential characteristics of the configurations from which CNN would explore.

\begin{table*}[!tbp]
	\centering
	\begin{tabular}{cccccccc}
        \hline
 initial condition &   dimension &Neural network    &$L=4$ &$L=8$   &$L=16 $  &$L=32$ &$L=48$  \\
        \hline
 single seed & $ (1+1)d $ & FCN  & 0.7403  & 0.7819 & 0.8183 &0.8373 & 0.8888   \\
fully occupied & $ (1+1)d $ &FCN  & 0.7698  &0.8675 &0.9137 &0.9577 &0.9683   \\
fully occupied & $ (2+1)d $ &FCN  & 0.9060  &0.9511 &0.9779  &0.9933 &0.9966   \\
fully occupied & $ (1+1)d $ &CNN  & 0.8236  &0.8856 &0.9261  &0.9622 &0.9743   \\
fully occupied & $ (2+1)d $ &CNN  & 0.9201  &0.9576 &0.9840   &0.9944  &0.9980       \\
        \hline
    \end{tabular}

\caption{
In supervised ML, test accuracies of different system sizes and dimensions ((1+1) and (2+1) dimensions) in two different neural networks (FCN and CNN).}
\label{table:15}
\end{table*}

\begin{figure*}[t]
%\begin{tabular}{cc}
\begin{tabular}{cc}
    \includegraphics[width=0.45\textwidth]{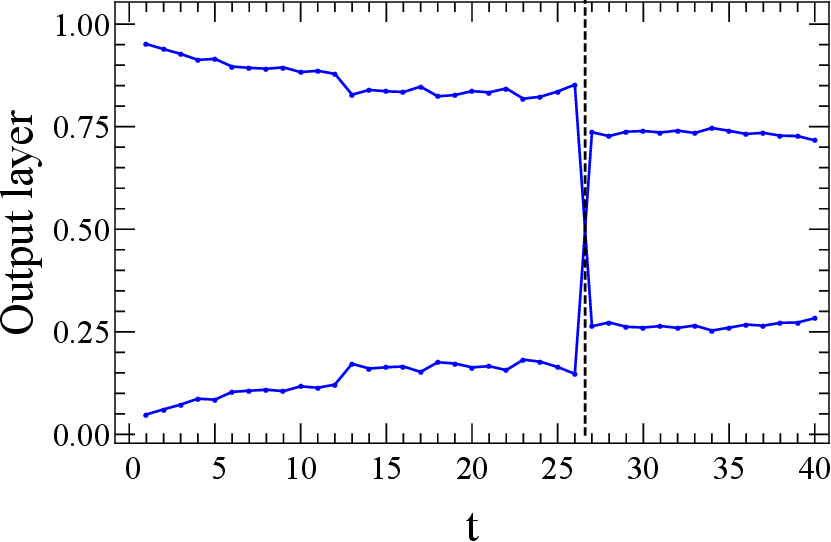} &
    $\qquad$\includegraphics[width=0.45\textwidth]{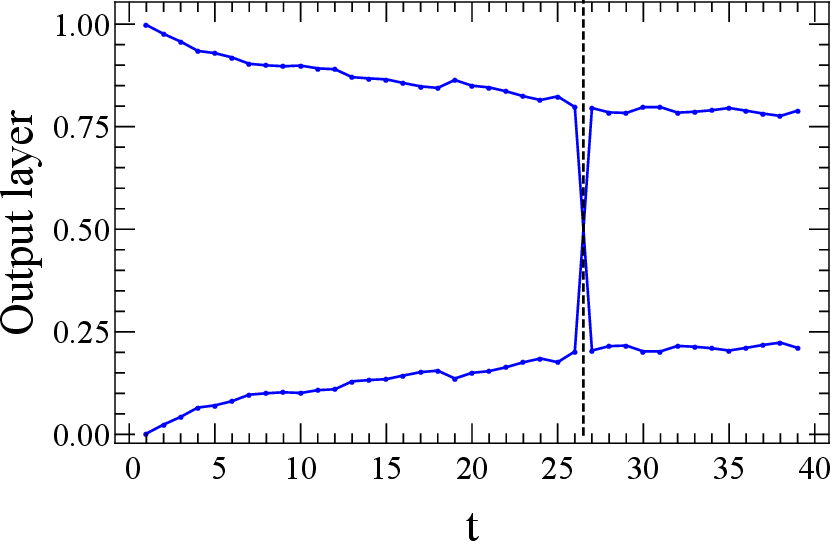} \\
    (a) & $\qquad$ (b)
\end{tabular}
\caption{ML characteristic time of the (1+1)-dimensional bond DP with FCN (a) and CNN (b), and the output layer averaged over a test set as a function of time step $t$, where the black dashed lines represent $t_{c}$. The system size is $L = 8$, and the time step is $40$.}
\label{841t}
\end{figure*}

\begin{figure*}[t]
\begin{tabular}{cc}
    \includegraphics[width=0.45\textwidth]{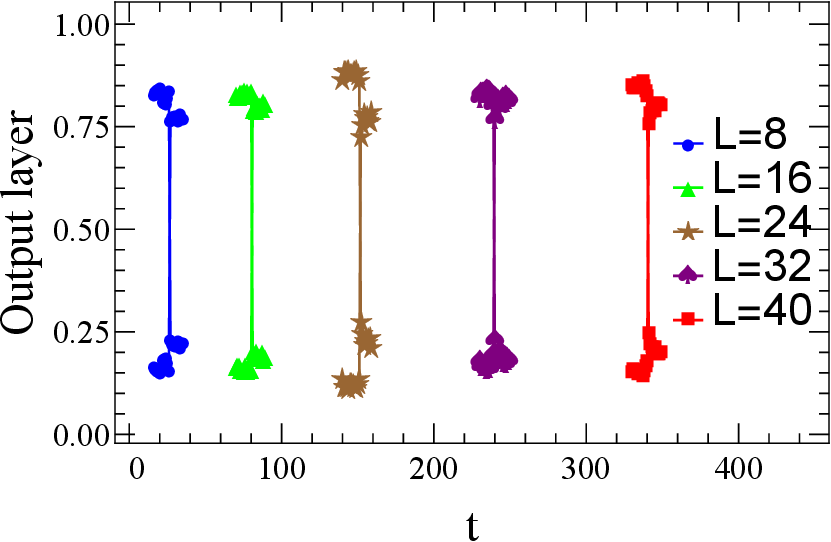} &
    $\qquad$\includegraphics[width=0.45\textwidth]{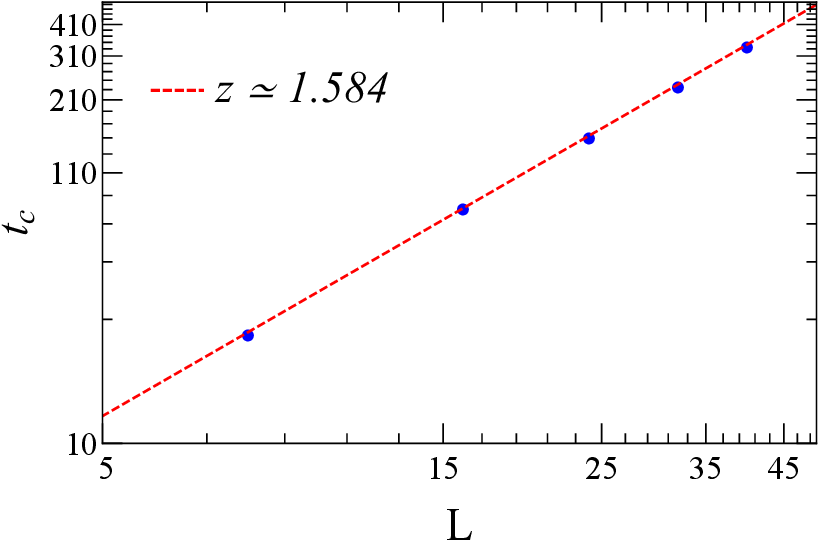} \\
    (a) & $\qquad$ (b)\\
    \includegraphics[width=0.45\textwidth]{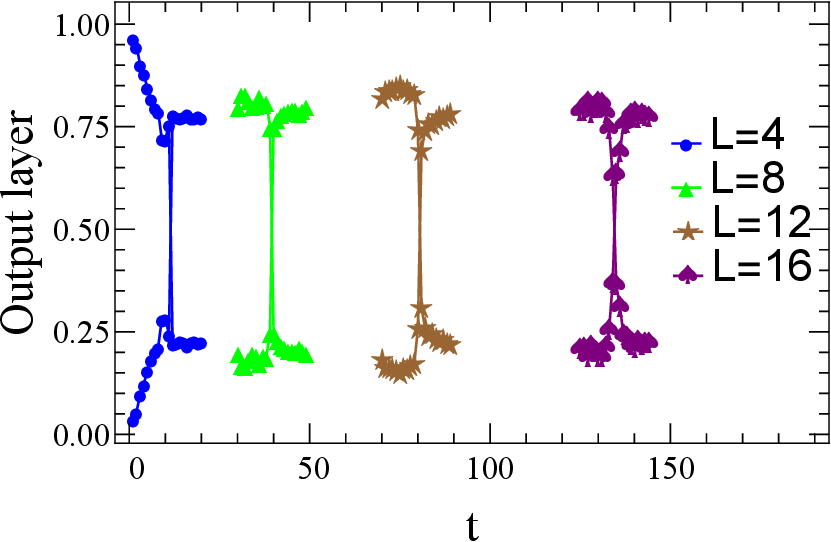} &
    $\qquad$ \includegraphics[width=0.45\textwidth]{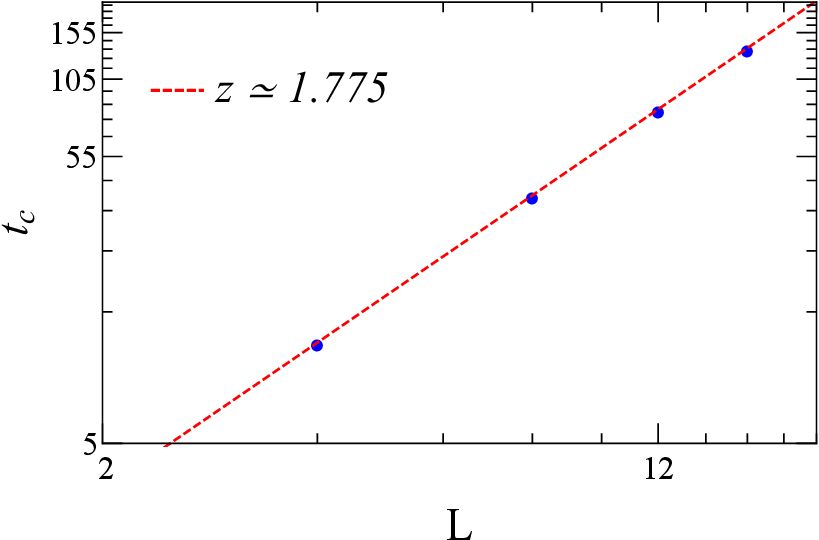} \\
    (c) & $\qquad$(d)
\end{tabular}
\caption{ML (1+1)-dimensional (top) and (2+1)-dimensional (bottom) bond DP. \textbf{a}, The output layer averaged over a test set as a function of time step $t$. \textbf{b}, The power-law fitting of characteristic time $t_{c}$ versus $L$ on a log-log scale. System sizes $L = 4, 8, 16, 24, 32, 40$ are represented by different colors, respectively.  \textbf{c}-\textbf{d}, Analogous data to \textbf{a}-\textbf{b}.}
\label{fig:time}
\end{figure*}

In (1+1)-dimensional bond DP, we have verified through ML if configurations are generated only between $0 \sim 0.4$ or $0.8 \sim 1$, this would lead to large uncertainty in predicating the critical threshold. It turns out that when using the supervised learning method in the DP model, the more dense labeled data we have in the training set, the more accurate the testing results we will achieve.

Besides, for ML of (1+1)-dimensional and (2+1)-dimensional bond DP, we also present their test accuracies in Table. \ref{table:15}. From Table. \ref{table:15}, it can be seen that with the same system size, the test accuracy of the (2+1)-dimensional bond DP is significantly higher than that of the (1+1)-dimensional one. When we use the same neural network to study the DP configuration, the absolute size of the higher dimensional system size will be larger, which could lead to more accurate results. This might be that in higher dimensions, particles have more degrees of freedom, and they diffuse better. In higher dimensions, even small ensemble averages can yield very accurate results. On the other hand, learning configurations of DP in the same dimension, CNN has a higher test accuracy than the FCN. It is mainly because, compared to the FCN, CNN shares only one parameter or weight during the training process.

\subsubsection{Optimal range and transformation effect}
So far, we have completed supervised ML for (1+1)-dimensional and (2+1)-dimensional bond DP. It seems that we could have some discussion over the optimal range of the bond probability and effect of the transformation of the input data.

\begin{figure}
\centering
\includegraphics[width=0.4\textwidth]{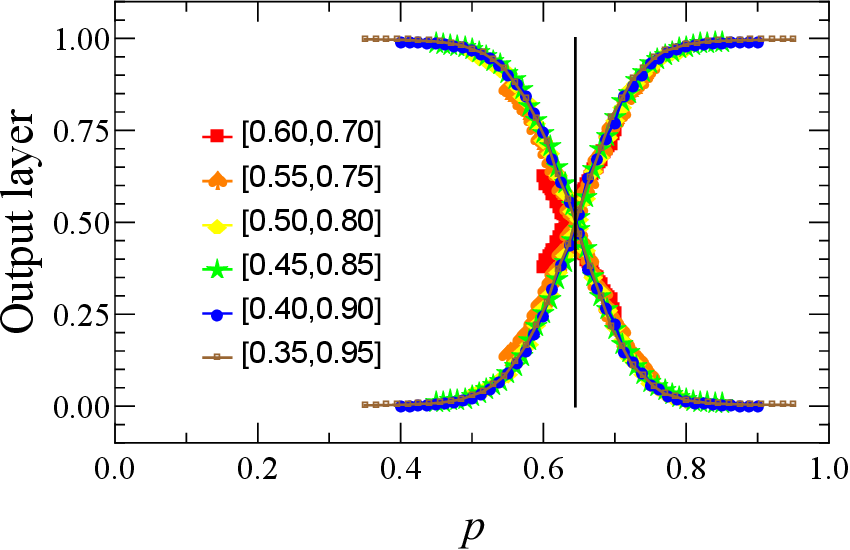}
\caption{ Supervised ML (1+1)-dimensional bond DP by FCN. The output layer, averaged over a test set, as a function of the bond probability $p$. The lattice size is $L = 16$.}
\label{range}
\end{figure}

In order to obtain the optimal sample value range for ML, we have conducted systematic and extensive experiments. There is unfortunately no theory about what range is optimal. The final bond probability range we select in (1+1)-dimensional DP is $[0.4,0.9]$. Here is how we make the choice. Table \ref{tablerange} displays the test accuracies of six different ranges. It can be seen that the last two ranges, $[0.4,0.9]$ and $[0.35,0.95]$, yield much higher accuracies than the rest of them. Fig. \ref{range} demonstrates the learning curves by the same six different ranges and still the last two have the best over-all performance. So combining Table \ref{tablerange} and Fig. \ref{range}, $[0.4,0.9]$ would be optimal with higher accuracy, better performance and less computational cost. It is interesting to notice that the red curve in Fig. \ref{range}, representing the range of $[0.6,0.7]$, which contains only 10 percent of the whole bond probability range $[0,1]$. But its test accuracy of is only $0.6913$ and the learning is very poor. This could be that neural networks do not have enough information to learn the trend of a series of configurations. Because close to the critical point, the configurations of the bond probability fluctuate greatly.

\begin{table*}[!tbp]
	\centering
	\begin{tabular}{cccccccc}
        \hline
range & [0.6,0.7] & [0.55,0.75]  & [0.50,0.80]  & [0.45,0.85]   &[0.40,0.90]  & [0.35,0.95]     \\
        \hline
test accuracy & 0.6913 & 0.8093  & 0.8686  & 0.9035 &0.9137  & 0.9251  \\

        \hline
    \end{tabular}

\caption{In supervised ML of (1+1)-dimensional bond DP, test accuracies of different bond probability ranges by FCN. The lattice size is $L= 16$.}
\label{tablerange}
\end{table*}

We have compared the learning curves for using FCN and CNN. In the case of (1+1) dimensions, the comparison of Fig. \ref{fig:res} (b) and Fig. \ref{mlcnn} (b) show that FCN and CNN can lead to the same prediction results of spatial correlation exponent. Fig. \ref{fig:res} (c) and Fig. \ref{mlcnn} (c) show that they also give the same values of critical points. The same comparisons in (2+1) dimensions have also been done and we find no significant difference of the critical points and critical exponent of interest. This indicates that the transformation on the input data, at least through FCN and CNN, has little effect on the final results.

\subsubsection{ML characteristic time $t_{c}$ of DP}
Conventionally, in equilibrium statistical mechanics, we can deal with statistical properties associated with static conditions. While in out-of-equilibrium cases, we have to process an explicit dynamical description, where time takes the form of an extra dimension. Non-equilibrium steady states do not satisfy detailed balance. DP displays a non-equilibrium phase transition from a fluctuating phase into an absorbing state, which we call an absorbing phase transition. Different from equilibrium phase transitions, absorbing phase transitions are characterised by at least three independent critical exponents $\beta$, $\nu_{\perp}$, $\nu_{\parallel}$ (or $z$).

Since learning the critical threshold of bond DP has been successful, it is natural to ask whether the supervised ML method is also applicable to the prediction of the characteristic time $t \sim L^{z}$ of the same model. Here we fix the given critical bond probability $p_{c} = 0.6447$ and chose different time steps to generate configurations. For example, for a lattice size of $L = 8$ and time steps of $t_{real} = 7$, the input data in the neural network is $l_{x} = L = 8$, and $l_{y} = t_{real} +1 = 8$.

In ML of the characteristic time $t_{c}$, for different system sizes, 2500 samples of configurations are selected for each time step. To avoid the data from being too large, for each system size, we only choose $20$ time steps adjacent to the characteristic time $t_{c}$ for ML. Nonetheless, we can achieve a test accuracy of at least $ 0.85 $. If we enlarge the interval of consecutive time steps, one can achieve a test accuracy above $0.90$.

Fig. \ref{841t} shows the results of FCN and CNN in supervised ML of characteristic time $t_{c}$. Here we only take FCN as an example. The intersection of the two curves gives the characteristic time. Configurations generated by time steps smaller than the characteristic time are labeled as "0", which indicates the system is still in an active state. Otherwise, configurations generated by time steps larger than the characteristic time are labeled as "1", which means that the system has reached an absorbing phase. We then take the intersections of the five different size results and present them on the $t_{c}$ versus $L$ in Fig. \ref{fig:time} (a). Finally, the power-law fitting yields the power exponent $z \simeq 1.58063$ in (1+1) dimensions, which is consistent with the theoretical value. On this basis, we conclude that the supervised ML can also be applied to the learning and prediction of the characteristic time $t_{c}$.

\section{Unsupervised ML of DP}
In dimensionality reduction for data visualization of phase transitions, many methods such as PCA \cite{wang2016discovering}, T-SNE \cite{zhang2019machine} and Autoencoder \cite{wetzel2017unsupervised,hu2017discovering} have been exploited. PCA is a linear dimensionality reduction method and is usually used only for processing linear data. Autoencoder, on the other hand, can be regarded as the generalization of PCA, which can learn nonlinear relations and has stronger ability to extract information. Autoencoder is based on network representation, so it can be used as a layer to construct deep learning networks. With proper dimensionality and sparse constraints, autoencoder can learn more interesting data projection than PCA and other methods. Here our purpose is mainly to extract latent features of the configuration data, so autoencoder is a suitable tool.

\begin{figure}
\centering
\includegraphics[width=0.4\textwidth]{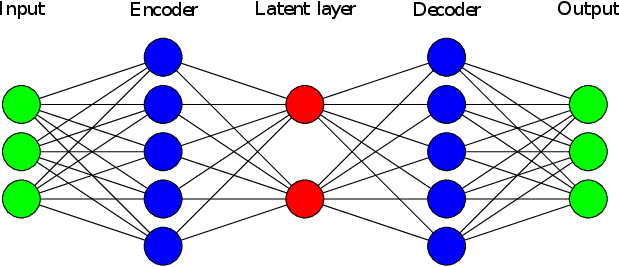}
\caption{ Neural network schematic structure of autoencoder.}
\label{autostruc}
\end{figure}

We regard a given configuration of (1+1)-dimensional bond DP as a two-dimensional image containing a time dimension. Through compression representations, the neural network non-linearly transforms and encodes input data, which is then transferred to the hidden neurons. Two hidden neurons are included in our setup of autoencoder, the structure schematic diagram of which is shown in Fig. \ref{autostruc}. Once a set of DP configurations generated by a series of bond probabilities are fed in, the autoencoder starts to train weights using methods like back propagation to return the target values equal to the input one. In this work, we choose a two-dimensional convolutional neural network as the encoding and decoding network. After passing through two convolutional layers and pooling layers in the encoding network, the data flow will be compressed by a fully connected layer into one or two hidden neurons. The decoding process is the inverse process of the encoding process, which can be regarded as a generative network. Here we focus on the dimensionality reduction process of encoding.

\begin{figure}
\centering
\includegraphics[width=0.4\textwidth]{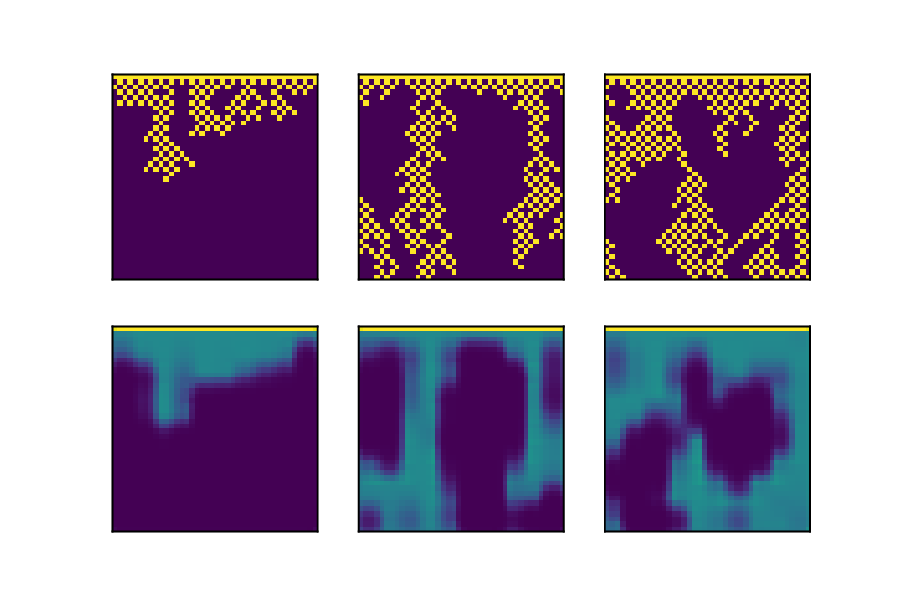}
\caption{  From the testing set, we choose three arbitrary (1+1)-d DP configurations (above) and reconstructions (below) using 200 hidden neurons. The original lattice size is $N = 40 \times 40$. }
\label{autoreconstruction}
\end{figure}

In the case of (1+1)-dimensional bond DP, the lattice size we choose to learn is $L=16$. According to $t_{c} \simeq L^{z}$, the corresponding characteristic time should be $t_{c} \simeq 80$. In bond DP, if $p > p_{c}$, the system could not reach an absorbing state until $T > t_{c}$. Artificially, we choose the time step $T=121$ which is larger than $T = 80$. By doing so, the neural network can receive more information about the configurations of DP.

In the encoder neural network, we assign 200 hidden neurons to the latent variable. In Fig. \ref{autoreconstruction}, we can see that the details of the raw DP configurations are lost during the reconstruction process. But the basic information of the input configuration is preserved in the latent variable. More Importantly, the general picture of many areas may still exist, so the latent variable is able to save information of the original configuration.

\begin{figure*}[!tbp]
%\begin{tabular}{cc}
%\begin{minipage}{0.48\linewidth}
%  \centerline{\includegraphics[width=6.0cm]{fewpoints.eps}}
%  \centerline{a}
%\end{minipage}
%\hfill
%\begin{minipage}{0.48\linewidth}
%  \centerline{\includegraphics[width=6.0cm]{autoencoder3.eps}}
%  \centerline{b}
%\end{minipage}
\begin{tabular}{cc}
    \includegraphics[width=0.5\textwidth]{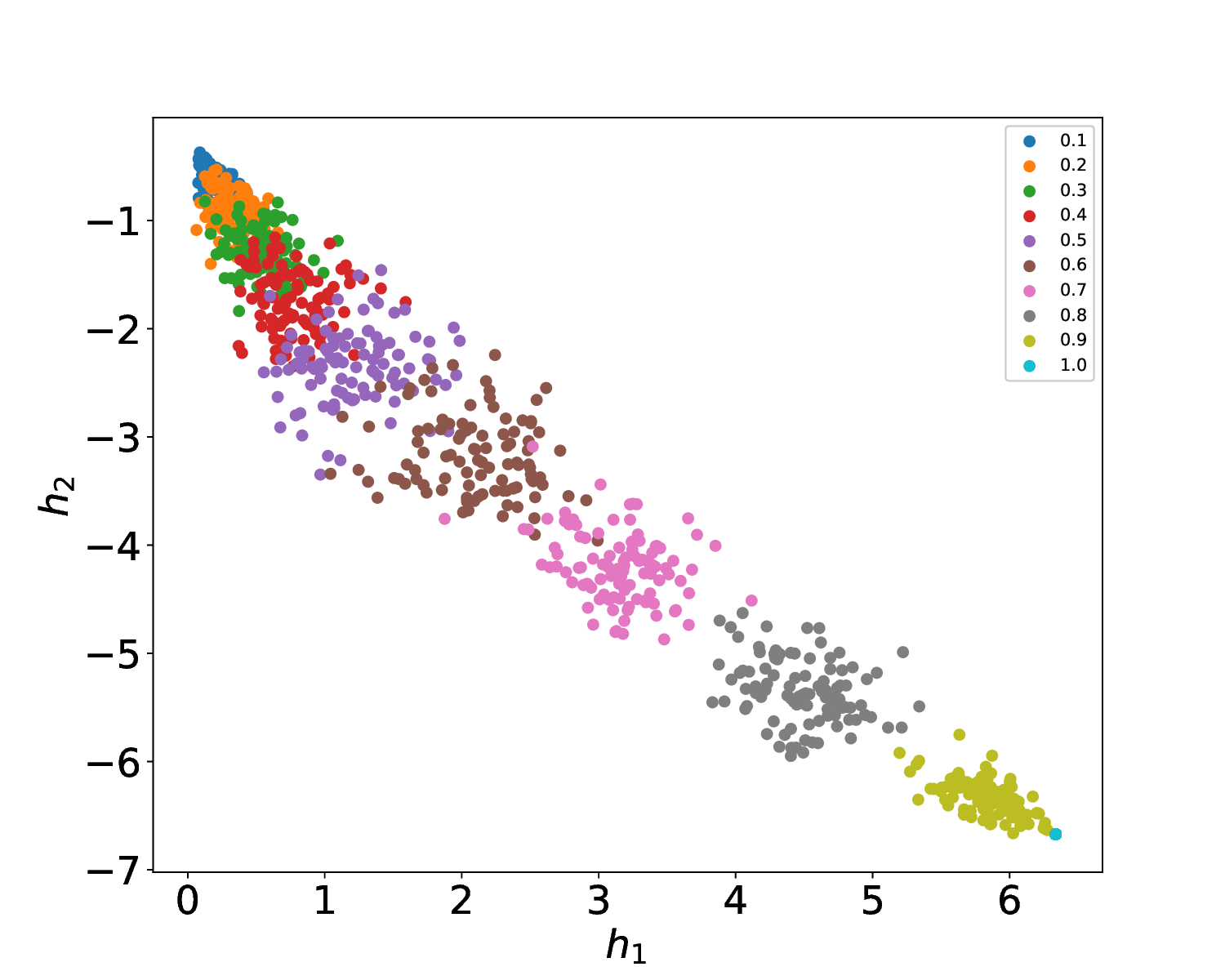} &
    $\qquad$\includegraphics[width=0.5\textwidth]{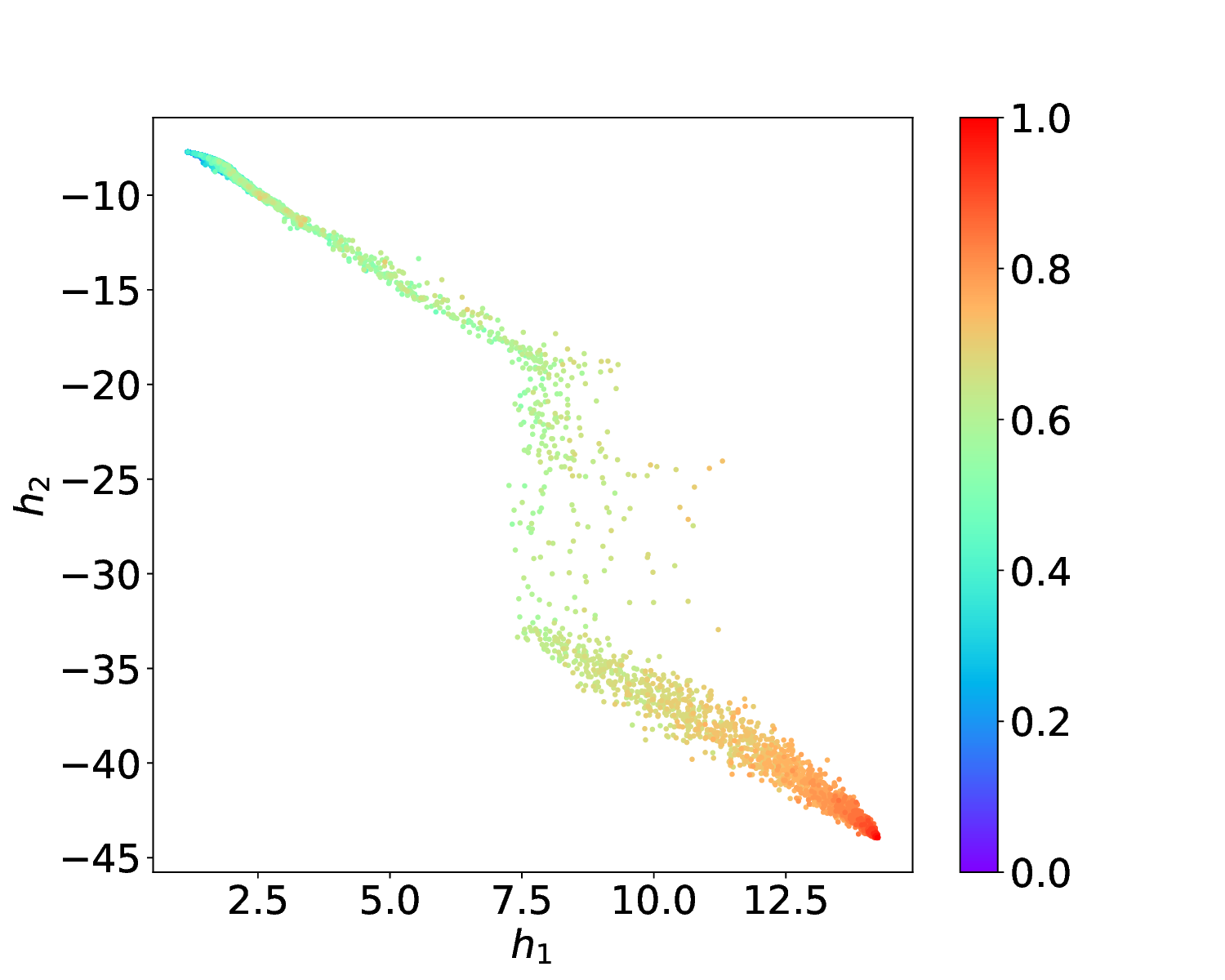} \\
    (a) & $\qquad$ (b)
\end{tabular}
%\end{tabular}
\caption{(1+1)-dimensional bond DP autoencoder results. Encoding of the raw DP configurations onto the plane of the two hidden neuron activations $(h_{1}, h_{2})$.}
\label{autoencodertwo}
\end{figure*}

First, we limit the autoencoder network to two hidden neurons and display the results in Fig. \ref{autoencodertwo}. Among them, the left part in Fig. \ref{autoencodertwo} represents the clusters of ten kinds of configurations generated by different bond probabilities, which are from $0.1 \sim 1$ with an interval of $0.1$, and each bond probability produces $100$ sets of configurations. We can clearly see that the configurations generated with the same probability are clustered in a close position. The right part of Fig. \ref{autoencodertwo} is the clustering diagram of $200$ configurations generated by $41$ different bond probabilities between $0$ and $1$ with equal intervals, from which we can find that obvious transition phenomenon occurs near a certain point. It is worth noting that, without changing any parameters of the neural network every time running the autoencoder neural network we may get a different cluster diagram. In particular, the horizontal and vertical coordinates of the $h_{1}$ versus $h_{2}$ visualization of the neural network would change too.

\begin{figure*}[t]
\begin{tabular}{cc}
    \includegraphics[width=0.45\textwidth]{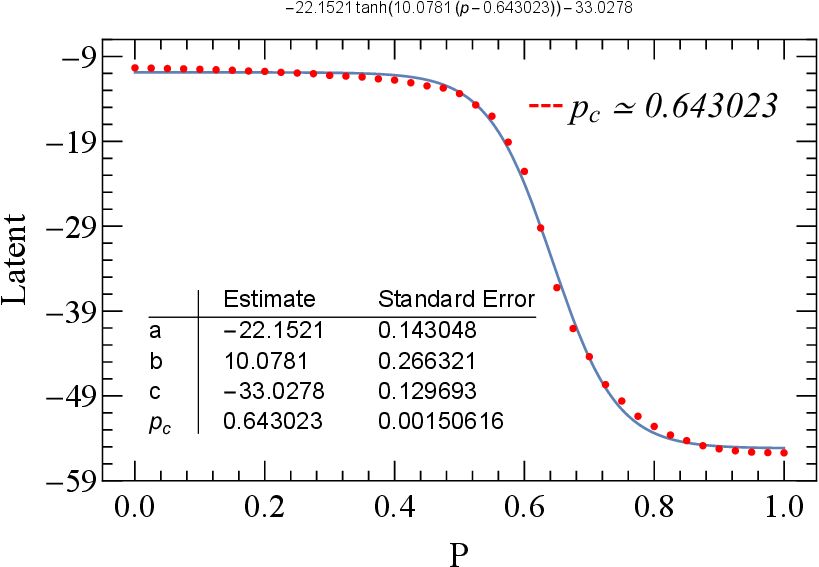} &
    $\qquad$\includegraphics[width=0.45\textwidth]{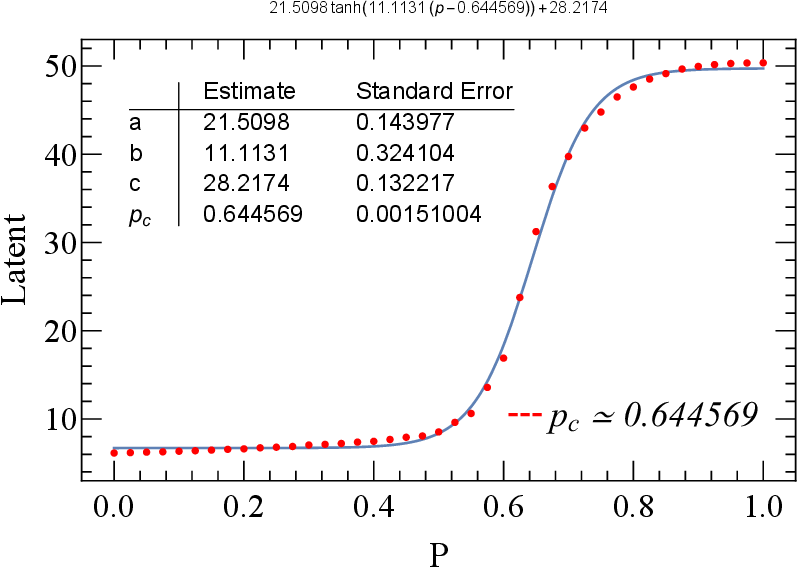} \\
    (a) & $\qquad$ (b)\\
    \includegraphics[width=0.45\textwidth]{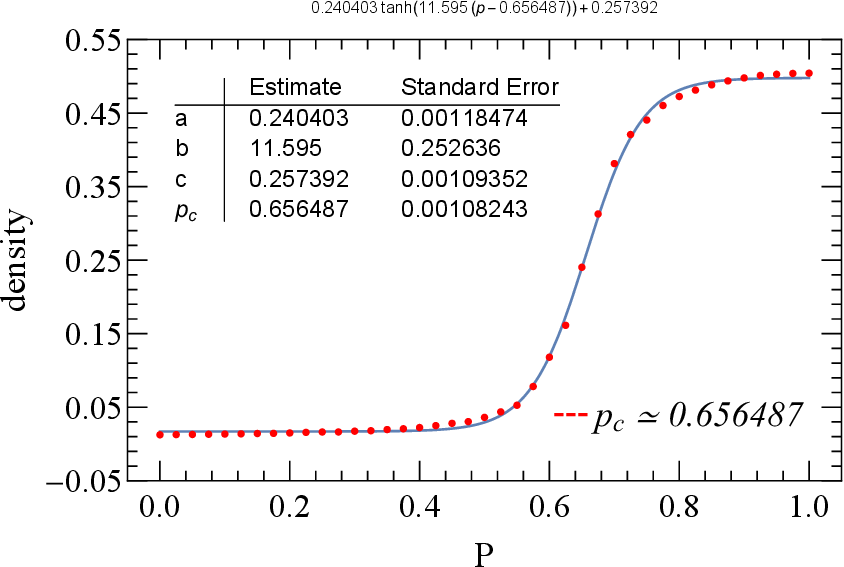} &
    $\qquad$ \includegraphics[width=0.45\textwidth]{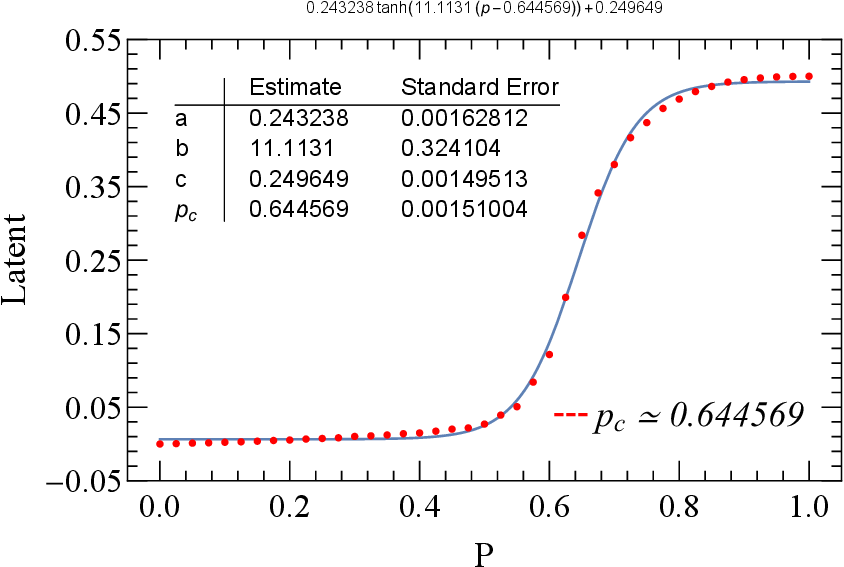} \\
    (c) & $\qquad$(d)
\end{tabular}
\caption{\textbf{a}-\textbf{b}, two similar autoencoder results of bond DP from one data set and convolutional autoencoder neural network. Encoding of the raw DP configurations using a single hidden neuron activation $Latent$ as a function of the bond probability. Each point of bond probability is averaged over $200$ testing samples.  \textbf{c}, (1+1)-dimensional bond DP Monte Carlo simulations, where the total time step is $t=121$, the lattice size is $N=16$, and the number of runs is $100$. \textbf{d}, with the same lattice size and time step, we get the normalization result of autoencoder in (1+1)-dimensional bond DP.}
\label{autoencoderone}
\end{figure*}

To find the transition point, we limit the latent layer of the two-dimensional convolutional autoencoder neural network to just one single hidden neuron whose results are displayed in the left part of Fig. \ref{autoencoderone}. Let the hyper-parameters of the autoencoder neural network remain unchanged, we continue to execute the autoencoder program and get another similar result, as shown in the right part of Fig. \ref{autoencoderone}. Take the left one as an example, red dots indicate that the single hidden neuron is regarded as a function of corresponding bond probabilities. To determine the critical point, we perform a non-linear fitting in the form of the hyperbolic tangent function $a \cdot tanh[b (p - p_ {c})] + c$ (the blue line is the fitting curve), and the jumping location $p_{c}$ is exactly what we are looking for. The prediction of the critical point of the left figure is $p_{c}\simeq 0.643(2)$, while the right one is $p_{c} \simeq 0.645(2)$. They are very close to the value given by the literature \cite{hinrichsen2000non} of (1+1)-dimensional bond DP, in which $p_{c} = 0.6447$.

%Nevertheless, we believe that more accurate results can be obtained, such as using more powerful computation resources, expanding training and testing samples, or executing finite-size scaling.

In order to capture the meaning of the information contained in the latent variables, we perform the following two studies. One is the reconstruction of the configurations shown in Fig. \ref{autoencoderone}, which shows that the latent variable contains basic information about the original configurations. On the other hand, we implemented another Monte Carlo simulation method for (1+1)-dimensional DP, and monitored the corresponding particle density for each bond probability. As with autoencoder, this Monte Carlo simulation corresponds to a lattice size of $L =16$ and a time step of $T = 121$, respectively. The result is shown in Fig. \ref{autoencoderone} (c), where the jumping point is the critical point, yielding approximately $p_{c} \simeq 0.656(1)$. Fig. \ref{autoencoderone} (d) is the same as Fig. \ref{autoencoderone} (b), and the only difference is that the y-axis of the former is rescaled by the maximum value. Intuitively, we find that the Monte Carlo simulation results and the autoencoder results have very similar trends. For the same bond probability, the particle density and the single latent variable have nearly the same curve. So we believe that the latent variable should be at least numerically proportional to the particle density, which is the order parameter. And from this point of view, it is not surprising that the transition point in the latent variable versus bond probability is simply the critical point.

The findings of this study indicate that the convolutional autoencoder neural network can accurately capture the phase transition point, and compress essential information of bond DP configurations into one neuron or two. 

We hope that this finding may shed some light on the study of non-equilibrium phase transitions. In general, the standard Monte Carlo simulations are costly, compared to the machine learning techniques. However, we notice that machine learning is not working for all the critical exponents. This is why we may consider combining the two techniques. ML can predict the critical points quickly and accurately whereas the Monte-Carlo simulations have its own advantages.

\section{Conclusion}
This paper mainly studied critical thresholds, spatial and temporal correlation exponents, and characteristic times of the bond DP model in both (1+1) and (2+1) dimensions by using artificial neural networks. Our conclusions are as follows.

Firstly, we successfully predicted critical thresholds and spatial and temporal correlation exponents of this model by using supervised ML. The results are in Tab. \ref{table2}. Through this, we found that supervised ML can give a prediction result very close to the literature. Additionally, with the same size, the test accuracy of the (2+1)-dimensional bond DP is significantly higher than that of the (1+1)-dimensional one. Even by reshaping the configurations of the complex system, neural networks can still learn its trend very well. And we also found that learning configurations of DP in the same dimension, CNN has a slightly higher test accuracy than FCN.

\begin{table}[ht]
	\centering
	\begin{tabular}{ccccc}
        \hline
                  & $(1+1)-d$ & $(1+1)-d$ & $(2+1)-d$ & $(2+1)-d$ \\
        \hline
      exponents   & literature & prediction & literature & prediction \\
        \hline
        $p_{c}$       & 0.6447    & $0.6408$ & $0.2873$  & $0.2855$\\
      $\nu_{\|}  $    & 1.733847  & $1.73 \pm 0.02$ & $1.295$ & $1.30 \pm 0.02$   \\

      $\nu_{\bot} = \nu_{\|}/ z$ & 1.096854  & $1.09 \pm 0.02$ & $0.73$ & $0.73 \pm 0.02$    \\
      $\beta=\delta \nu_{\|}$  & 0.276486  & 0.2728 &                   \\
        \hline
   \end{tabular}

\caption{Prediction values of DP in supervised ML. }
\label{table2}
\end{table}

Second, regarding time as an additional dimension of non-equilibrium phase transitions, we then studied the characteristic time $t_{c}$ of the bond DP from a fully occupied initial state to an absorbing state by supervised ML. As we can see, the predictions of the two neurons of the output layer by using FCN do not follow smooth curves, which implies that the system is greatly affected by finite-size effects and fluctuations. However, when CNN is employed, we can have much better curves. After training the configurations generated by different time steps, the neural network will identify whether the system has reached an absorbing state or not, as a new similar configuration is fed in. Then, the intersection of two output neuron curves is the predicted characteristic time.

Third, the (1+1)-dimensional bond DP is learned by using the unsupervised autoencoder method. We not only successfully made a clustering of its configurations generated by different bond probabilities but also obtained an estimation of the critical point.

When handling the problem of DP with Monte Carlo simulations, we need a sufficiently large lattice size $L$ and time steps $t$ to obtain a relatively accurate critical threshold of $p_{c}$. However, ML works for much smaller systems. The research of phase transitions in statistical physics is greatly benefiting from ML algorithms. ML can not only learn and predict the critical point but also estimate critical exponents of correlation exponents when the system approaches the critical point. Analogous to the Ising model in equilibrium phase transitions, the raw configurations of DP in non-equilibrium phase transitions can also be learned and predicted by neural networks.

In summary, in this paper we argued that ML techniques can be applied to non-equilibrium phase transitions. It is optimistic that with the continuous advancement of computer hardware technology and ML algorithms, plenty of ML techniques will be extended to solving problems of non-equilibrium phase transitions in statistical physics.

%\FloatBarrier
\section{Acknowledgements}
We wish to thank Juan Carrasquilla and Wanzhou Zhang for the helpful computer programs. We thank the two referees for their constructive suggestions and comments. This work was supported in part by the Fundamental Research Funds for the Central Universities, China (Grant No. CCNU19QN029), the National Natural Science Foundation of China (Grant No. 11505071, 61702207 and 61873104), and the 111 Project 2.0, with Grant No. BP0820038.

\bibliographystyle{apsrev4-2}
\bibliography{transfer_learningref}

\end{document}